\documentclass[aps,pre,twocolumn,superscriptaddress]{revtex4-1}
\usepackage[utf8]{inputenc}
\usepackage{amsmath}
\usepackage{amssymb}
\usepackage{epsfig}
\usepackage{graphicx,url}
\usepackage{bm}
\usepackage{mathrsfs}
\usepackage{tikz}
\usepackage{algpseudocode}
\usepackage{subcaption}
\usepackage{comment}
\usepackage{enumitem}

\begin{document}

\title{Alignment interaction and band formation in assemblies of auto-chemorepulsive walkers}

\author{Hugues Meyer}
\affiliation{Department of Theoretical Physics \& Center for Biophysics, Saarland University, 66123 Saarbrücken, Germany}
\author{Heiko Rieger}
\affiliation{Department of Theoretical Physics \& Center for Biophysics, Saarland University, 66123 Saarbrücken, Germany}
\affiliation{INM – Leibniz Institute for New Materials, Campus D2 2, 66123 Saarbrücken, Germany}

\date{\today}

\begin{abstract}
	Chemotaxis, i.e. motion generated by chemical gradients, is a motility mode shared by many living species that has been developed by evolution to optimize certain biological processes such as foraging or immune response. In particular, auto-chemotaxis refers to chemotaxis mediated by a cue produced by the chemotactic particle itself. Here, we investigate the collective behavior of auto-chemotactic particles that are repelled by the cue and therefore migrate preferentially towards low-concentration regions. To this end, we introduce a lattice model inspired by the {\it true self-avoiding walk} which reduces to the Keller-Segels model in the continuous limit, for which we describe the rich phase behavior. We first rationalize a the chemically-mediated alignment interaction between walkers in the limit of stationary concentration fields, and then describe the various large-scale structures that can spontaneously form and the conditions for them to emerge, among which we find stable bands traveling at constant speed in the direction transverse to the band. 
\end{abstract}

\maketitle

\section{Introduction}

Chemotaxis refers to motion induced by the presence of one or multiple chemical compounds. This feature is observed in nature in many forms, mostly in cells \cite{Bagorda2008, Roussos2011} and bacteria \cite{Adler1975, Wadhams2004} and is often used for particles to optimize search processes such as foraging, as the {\it smell} of the targeted object can be detected and used as a guide. Among the many forms of chemotaxis, auto-chemotaxis occurs when a particle emits itself the chemical cue to which it is sensitive. As a well-known example, many species of ants deposit pheromones along their path when foraging, which they then use to find their way back \cite{Traniello1989}. We can also mention the phenomenon of neutrophil swarming \cite{Kienle2016}, occuring when a neutrophil emits a chemo-attractant to recruit other immune cells in its vicinity for assistance in the killing process. On the other hand, auto-chemorepulsion, i.e. when particles are repelled by the chemical they produce, is also observed in nature. {\it Dictyostelium discoideum}, an amoebian species, is e.g. a example of auto-chemorepulsive organism that has been studied experimentally for many years now \cite{Song2006,Willard2006, Annesley2009}. 

Chemotaxis was first quantitatively modeled by Keller and Segel who proposed in 1971 field equations describing the time-evolution of both the chemical concentration field and the particle density field \cite{Keller1971}. These equations in their most general form account for various processes inherent to chemotaxis, namely the emission of the cue by the particles, the diffusion and degradation of the cue, the diffusion of the density field and its advection due to chemical concentration gradients. The Keller-Segel model has been extensively studied and generalizations have been proposed over the past decades, such as a fractional form of the model \cite{Horstmann2003,Hillen2009,Arumugam2021,Escudero2006}. More recently, various particle models for chemotaxis and auto-chemotaxis have been introduced and molecular simulations of such models have been performed. Most of the recent studies focused their attention on chemo-attraction for which self-organized structures were discovered and thoroughly described \cite{Taktikos2011, Pohl2014, Jin2017, Liebchen2020, Murugan2022, Mokhtari2022, Kolk2022, Traverso2022}. However, less has been reported on auto-chemorepulsion. From the pure point of view of statistical physics, auto-chemorepulsive walks are interesting as they can be seen as random walks with memory since particles tend not to visit twice regions that they have previously visited, provided that the chemical cue has not diffused away. Such type of non-markovian walks where particles have some memory of their previous locations has motivated some recent research \cite{Schutz2004, Boyer2014, Meyer2021, Barbier2022}, with the aim of characterizing their statistical properties such as first-passage times or record statistics. Understanding the collective behavior of auto-chemorepulsive particles and how they self-organize is therefore a missing piece of this research field at the interface of physics and biology. 

In this paper, we introduce a lattice model for auto-chemorepulsive walkers. It is inspired by the {\it true self-avoiding walk} \cite{Amit1983}, where a walker jumps to a neighboring site with a probability weighted by the number of times it has already visited it, to which we add a diffusion step. The model reduces to the Keller-Segel model in the continuum limit, and is controlled by two main parameters. First, the concentration diffusion constant, which essentially acts as a the memory of the walk; and the coupling of a walker to the concentration field, which controls the persistence of the walk. The main result that we present in this paper is the formation of bands of particles, traveling at constant speed in the direction transverse to the band. Similar bands have not only been observed in experimental bacterial systems for several decades \cite{Adler1966, Adler1967, Dahlquist1972, Mazzag2003}, but also in various models of active systems \cite{Peruani2011, Solon2015, Martin2018, Mangeat2020, Chatterjee2020} and very often results from an alignment interaction, as in the well-known Vicsek model \cite{Vicsek1995}. In the case of auto-chemorepulsive particles, the alignment interaction is mediated by the concentration field and depends on the diffusion constant of the chemical cue in a non-trivial way. 

The paper is organized as follows. We first introduce the lattice model and show its connection to the Keller-Segel model. In a second section, we discuss the phenomenology of the interaction between the walker and the field, by first characterizing the self-interaction between a walker and the cue it has produced itself, and then by computing alignment probabilities of two particles going in different directions. In a third part, we show results of numerical simulations where we identify three different phases that break the directional symmetry. The conditions for these phases to form are presented in a phase diagram and discussed in details. We conclude the paper by discussing the implications of our results.

\section{The model}

Recently, computational models for auto-chemotactic particles have been introudced in the literature. For instance, the model considered in \cite{Pohl2014, Stark2018} represents auto-chemotactic particles as active Brownian particles which experience a translational force as well as a torque proportional to the concentration gradient of the chemical cue. This is coupled to a diffusion equation for the concentration field which includes a source term located at the positions of each particle. While this type of detailed continuous approach allows to characterize realistic systems, it has the drawback of being computationally expensive as the resolution needed for solving the diffusion equation can be prohibitive.  

In order to avoid computationally expensive simulations, we propose a minimal lattice model for autochemotactic particles that contain three main ingredients : (i) particles emit a chemical cue along their path, (ii) they migrate preferentially to regions of low chemical concentration, (iii) the cue diffuses according to normal diffusion. To this end, we consider a $d$-dimensional lattice on which a concentration field $c$ is defined, its value on a site $i$ at time $t$ being noted $c_i(t)$. $N_w$ walkers are placed on sites of the lattice and we note $\rho_i$ the number of particles on lattice site $i$. The total walker density is noted $\varrho$. The time-evolution of the system must couple to the diffusion of the concentration field, the motion of the walkers on the lattice, controlled by the local value of the concentration field, and their production of the chemical cue. We therefore introduce the following algorithm to evolve the system over one time step from time $t$ to $t+1$:
\begin{enumerate}
	\item Generate a random permutation $\mathcal{P}$ of the set $\left\{0,\cdots,N_w-1 \right\}$

	\item For $m = 0 , \cdots, N_w-1$
	\begin{enumerate}
		\item Move walker $l=\mathcal{P}_m$ from its current site $i$ to a neighboring site $j$ with probability
		\begin{equation}
			p_{i\to j} = \frac{e^{-\beta c_{j} } }{\sum_{k\in\mathcal{N}_i}e^{-\beta c_k }}
		\end{equation}
		where $\mathcal{N}_i$ is the set of neighboring sites of $i$. 
		\item Update $c_j \gets c_j + h$
	\end{enumerate}
	\item Integrate on a time interval of length $1$ the following diffusion equation 
	\begin{equation}
		\label{eq:chemical_diffusion}
		\frac{d c_i(t)}{dt} = D_c\sum_{j\in \mathcal{N}_i}\left[c_{j}(t)-c_i(t)\right] 
	\end{equation}
\end{enumerate} 
The model is governed by three main parameters:
\begin{itemize}
	\item the deposition rate $h$. We define it as a unit of concentration per unit time and therefore use $h=1$ throughout this study. 
	\item the diffusion constant $D_c$. For $D_c= 0$, any amount of chemical deposited on a site remains there for ever, and the model becomes the well-known true self-avoiding walk (TSAW).
	\item the chemotactic coupling strength $\beta$. For $\beta=0$, the walkers are not sensitive to the concentration field and jump to any neighboring site with the same probability. On the other hand, for $\beta\to\infty$ the walkers always jump to the neighboring site with the lowest concentration.
\end{itemize}

For the rest of this paper, we will focus on a 2-dimensional system and we will use the following notations. $s_l$ refers to the site occupied by the walker $l$, whose coordinates are $(x_l, y_l)$. The current direction of this walker is $\mathbf{v}_l$ and is a vector that can only be $(0,1), (0,-1), (1,0)$ or $(-1,0)$. We also introduce the directed density field $\rho_\mathbf{v}$ where ${\rho_\mathbf{v}}_i$ is the number of walkers going in the direction $\mathbf{v}$ located at the site $i$. Given $(x,y)$ the coordinates of the site $i$, we equivalently use the notations $\rho(x,y) = \rho_i, \rho_\mathbf{v}(x,y)= {\rho_\mathbf{v}}_i$ as well as $c(x,y) = c_i$ for the chemical concentration field.

A natural question that arises when working with a discrete lattice model is its continuum limit in which the lattice constant is smaller than any other length scale involved in the problem. To answer this question, let us first consider the continuuum limit of the 1-dimensional case.

Let us define the time-dependent continuous concentration field $C(x,t)$ and density field $\rho(x,t)$, such that $c_i(t) = C(ia, t)$ and $n_i(t) = a \rho(ia, t)$, where $a$ is the size of a lattice site and $n_i$ is the number of walker on site $i$.

The partial differential equation for the concentration field can be derived from the discrete equation (\ref{eq:chemical_diffusion}). A standard finite difference coefficient analysis for the spatial dependence yields, for $a\to 0$:
\begin{equation}
\label{eq:continuous_chemical}
\frac{\partial c(x,t)}{\partial t} = D_c \frac{\partial ^2c(x,t)}{\partial x^2} + h a\sum_{k} \rho (x,t) \delta(t-k\delta t) 
\end{equation}
where $\delta t$ is the time step in the discrete case. At this point, the source term takes a complicated form as the walkers add the chemical cue at once whenever they jump. However, for $\delta t \to 0$ we have 
\begin{equation}
	 \int_{t}^{t + \delta t} dt' \rho (x,t') \delta(t-k\delta t')  \simeq  \rho (x,t) 
\end{equation}
such that we can approximate the equation as 
\begin{equation}
\frac{\partial c(x,t)}{\partial t} = D_c \frac{\partial ^2c(x,t)}{\partial x^2} + h  a \rho (x,t)  
\end{equation}

The second equation governs the time-evolution of the density field. To derive it from the discrete case, we consider the site $i$, at position $x$. The number of particles at this site is noted $N(x,t)$. The rules of our discrete autochemotactic walk imply the following master equation
\begin{align}
\rho(x,t+\delta t) =& \frac{\rho(x-a,t)}{1+e^{-\beta\left(c(x,t)-c(x-2a,t) \right)}} \nonumber \\
& + \frac{\rho(x+a,t)}{1+e^{-\beta\left(c(x,t)-c(x+2a,t) \right)}}
\end{align}
where $a$ is the size of a lattice site. Expanding for $a \ll 1$ and $\delta t \ll 1$ yields
\begin{equation}
\label{eq:continus_density}
\frac{\partial \rho(x,t)}{\partial t}  = D_p \left[ \frac{\partial^2\rho(x,t)}{\partial x^2} + 2\beta \frac{\partial}{\partial x}\left(\rho(x,t) \frac{\partial c(x,t)}{\partial x} \right) \right] 
\end{equation}
where $D_p = av_0/2$ is the particle diffusion constant and $v_0=a/\Delta t$ is the velocity of the particles. This is a diffusion equation with a drift term proportional to the gradient of the concentration field. The set of equations (\ref{eq:continuous_chemical}) and (\ref{eq:continus_density}) is an instance of the Keller-Segel model which accounts in its most general form for multiple processes. The same analysis can be made in higher dimension and lead to the same conclusion. Note that the drift velocity in (7) is $v=2\beta dc/dx$ and is therefore negative when the concentration gradient is positive, which implies chemo-repulsion. The original Keller-Segel model was devised to describe chemo-attraction and therefore contains a minus sign on the right-hand side. The Keller-Segel model has been studied in many publications since its introduction in 1971 \cite{Keller1971}, but the case of chemorepulsion has not been extensively discussed \cite{Liebchen2015}. One should therefore keep in mind that the results we will show for the lattice model we are using can be extended to the Keller-Segels model of auto-chemorepulsion if the lattice constant is small enough, which practically corresponds to large diffusion constants $D_c$.

\section{Phenomenology of the field-mediated interaction}

The only information that a walker uses to decide which site to jump to at the next step is the local value of the concentration field produced by itself and other walkers. We discuss here the phenomenology of this field-mediated interaction. In general, two walkers $1$ and $2$ located at positions $\mathbf{r}_1$ and $\mathbf{r}_2$ which have respectively produced a concentration field $c_1$ and $c_2$, are likely to align if $\left[\mathbf{\nabla}c_1(\mathbf{r}_1)+\mathbf{\nabla}c_2(\mathbf{r}_1)\right]\cdot\left[\mathbf{\nabla}c_1(\mathbf{r}_2)+\mathbf{\nabla}c_2(\mathbf{r}_2)\right]$ is large. In practice, the fields $c_{1,2}$ depend on the current position of the particle and on the complete trajectories of all walkers. Here we consider a few simplified scenarios to understand emerging effective interactions.

\begin{figure}
	\begin{center}
		\includegraphics[width=\linewidth]{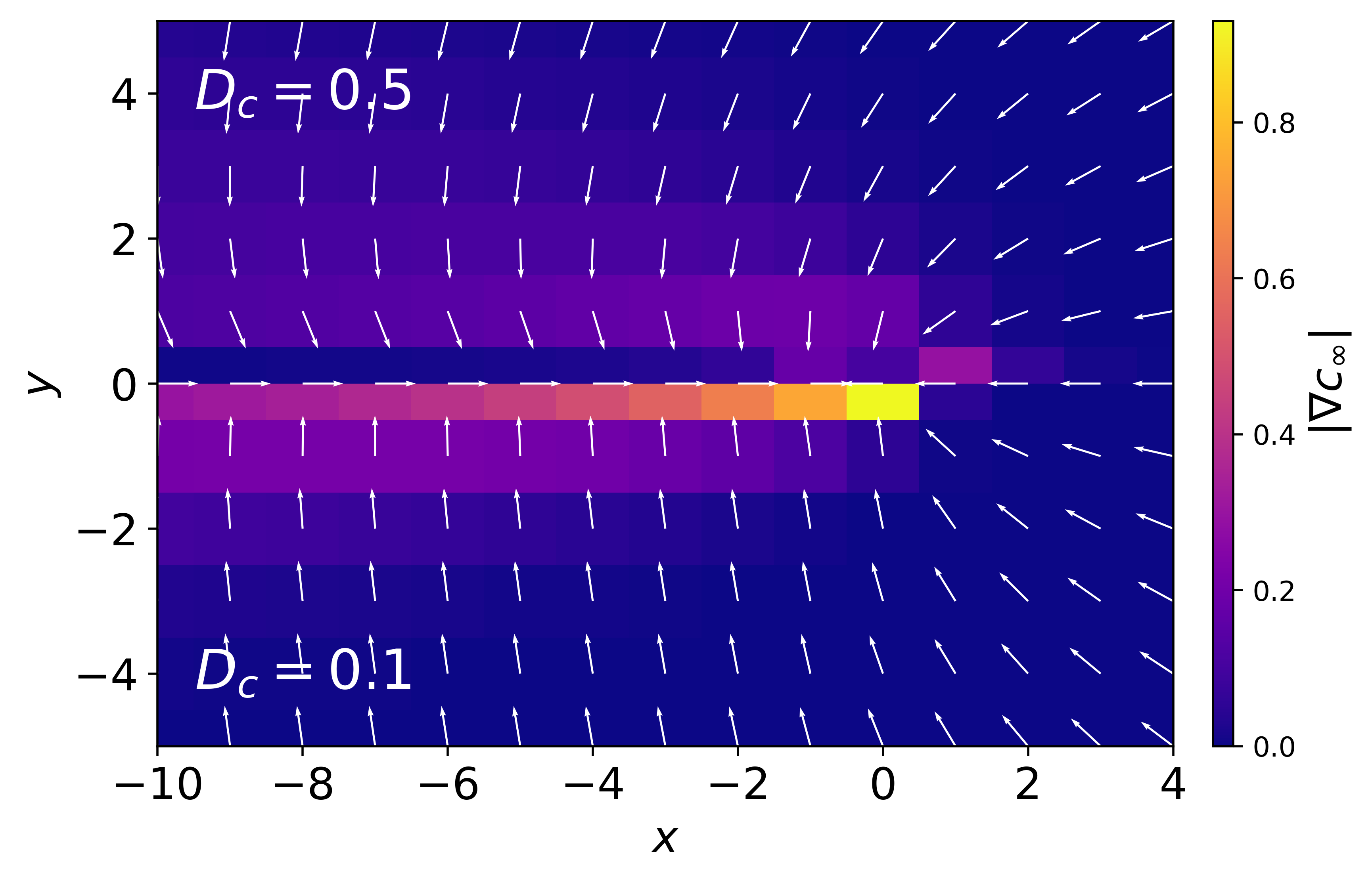}
		\caption{Concentration gradient $\nabla c_\infty$, where the color indicates its norm and the arrow its direction. The lower-half plan ($y>0$) corresponds to $D_c=0.1$ while the upper-half to $D_c=0.5$.}
		\label{fig:gradients}
	\end{center}
\end{figure}

As one walker follows on a straight line in the direction $\mathbf{u}$, the concentration field that it has produced reaches a stationary profile centered around it in the long-time limit. Let $c_\infty(\mathbf{u}; \mathbf{r})$ be this profile, where $\mathbf{r}$ indicates the vectorial distance to the site occupied by the walker. Symmetry under rotation imposes that  $c_\infty(\mathbf{R}_\theta \mathbf{u}; \mathbf{r}) = c_\infty(\mathbf{u}; \mathbf{R}^{-1}_\theta\mathbf{r})$ where $\mathbf{R}_\theta$ is the rotation matrix of angle $\theta$. We therefore omit the dependence on $\mathbf{u}$ and use the convention $c_\infty(\mathbf{r}) = c_\infty(\mathbf{e}_x, \mathbf{r})$. We show in figure \ref{fig:gradients} the gradient $\mathbf{\nabla} c_\infty$ for $D_c=0.1$ and $D_c=0.5$, defined as $\mathbf{\nabla} c_\infty(x,y)=\frac{1}{2}(c_\infty(x+1,y)-c_\infty(x-1,y), c_\infty(x,y+1)-c_\infty(x,y-1))$, whose additive inverse can be interpreted as the effective force exerted by the cue produced by a walker on an other walker located at a distance $\mathbf{r}$. As expected the concentration gradient is oriented mostly orthogonally to the trail left behind the walker, with a small component in the direction of the trail. In addition, the gradient at the location of the walker is directed along the trail, in the direction of the walk. This structure impacts the interaction of a walker with its own field but also the alignment interaction of walkers depending on their relative directions, as discussed next. 

\subsection{Self-interaction}

\begin{figure}
	\begin{center}
		\includegraphics[width=\linewidth]{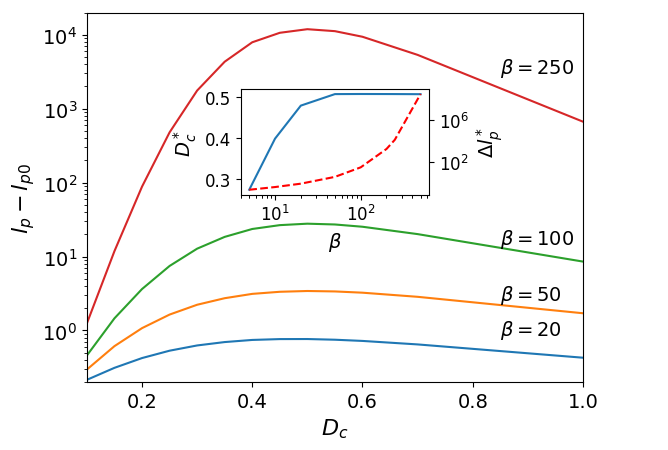}
	\end{center}
	\caption{Persistence length $l_p$ as a function $D_c$ and various values of $\beta$, where we have subtracted the value ${l_p}_0 = 1/3$ reached for $\beta = 0$. The inset shows the value of $D_c$ that maximizes the persistence length as a function of $\beta$ (solid line), together with the corresponding value of the persistence length $\Delta l_p^* =l_p^* - {l_p}_0$ (dashed line).}
	\label{fig:single_lp}
\end{figure}

Before considering the effective interaction between two walkers, we characterize the force exerted by the concentration field produced by a walker on the walker itself. To do this, we consider the stationary field $c_\infty(\mathbf{r})$. As this field results from an infinite straight walk, the value of the field on the site that has just been visited by the walker is necessarily higher than the left and right sites and even higher than on the forward site. For $\beta>0$, the probability for the walker to chose the forward site will therefore be the highest such that it will preferentially continue in the same direction. The concentration field therefore effectively acts as an aligning force that hinders the walker to turn. This effect can be quantified by a persistence length $l_p$ as the mean number of steps continued along the same direction, starting from the stationary field $c_\infty$, reading
\begin{equation}
l_p = \sum_{k=0}^{\infty} k a_{\rightarrow}^k \left(1-a_{\rightarrow}\right)= \frac{1}{a^{-1}_{\rightarrow} - 1}
\end{equation} 
where $a_{\rightarrow}$ is the probability for the walker to continue in the same direction at the next time step and is given by 
\begin{equation}
a^{-1}_{\rightarrow} = 1 + \sum_{\mathbf{u}\neq \mathbf{e}_x} e^{-\beta \left( c_\infty(\mathbf{u})-c_\infty(\mathbf{e}_x)\right) }
\end{equation}
Noting $\Delta c_\infty^\perp = c_\infty(\mathbf{e}_y)-c_\infty(\mathbf{e}_x)$ and $\Delta c_\infty^\parallel = c_\infty(-\mathbf{e}_x)-c_\infty(\mathbf{e}_x)$ we obtain
\begin{equation}
l_p = \left[2e^{-\beta \Delta c_\infty^\perp} + e^{-\beta \Delta c_\infty^\parallel}\right]^{-1}
\end{equation} 
Because of mass conservation, both $\Delta c_\infty^\perp$ and $\Delta c_\infty^\parallel$ can not be arbitrarily large simultaneously.  We show in figure \ref{fig:single_lp} the persistence length as a function of $D_c$ for various values of $\beta$. It reaches a $\beta$-dependent maximum which results from two competing effects. First, for $D_c \to 0$, the cue diffuses so slowly that $\Delta c_\infty^\parallel$ is large but $\Delta c_\infty^\perp$ is very low, such that a turn to the right or left is probable. On the other extreme for $D_c \to \infty$, the cue has diffused so fast in one time step that both $\Delta c_\infty^\parallel$ and $\Delta c_\infty^\perp$ are low. This again implies that a turn (and even a reverse jump) is easy. In between these two limits, there exists a region where the cue has diffused sufficiently for $\Delta c_\infty^\perp$ to be non-zero but has not diffused enough for $\Delta c_\infty^\parallel$ to be insignificant. In such a situation, the forward site will be highly favored for the walker to jump to and the persistence length will be substantially increased.

We conclude that the self-produced concentration field acts as an aligning force whose strength is maximized for a certain value of the diffusion constant $D_c$.
 
\subsection{Aligning two walkers}

Now, let us consider two walkers going in directions $\mathbf{u}_1$ and $\mathbf{u}_2$ for a long enough time such that the field they produce is the stationary one. As they approach each other, they will sense more and more the field produced by the other walker. We therefore ask the question: given their mutual distance, what is the probability that they will align their directions ? To formalize the answer, we need to consider the combined field $c^{(2)}_{\infty}(\theta, \mathbf{\Delta}; \mathbf{r}) = c_\infty(\mathbf{r}) + c_\infty(R_\theta \left(\mathbf{r}-\mathbf{\Delta}\right))$. In other words, $c^{(2)}_\infty$ is the field resulting from the infinite straight walks of two walkers oriented with and angle $\theta$ and which are separated by a vector $\mathbf{\Delta}$, with the origin placed at the position of the first walker. 

\subsubsection{$\theta = \pi/2$}
\begin{figure}
	\begin{center}
		\includegraphics[width=\linewidth]{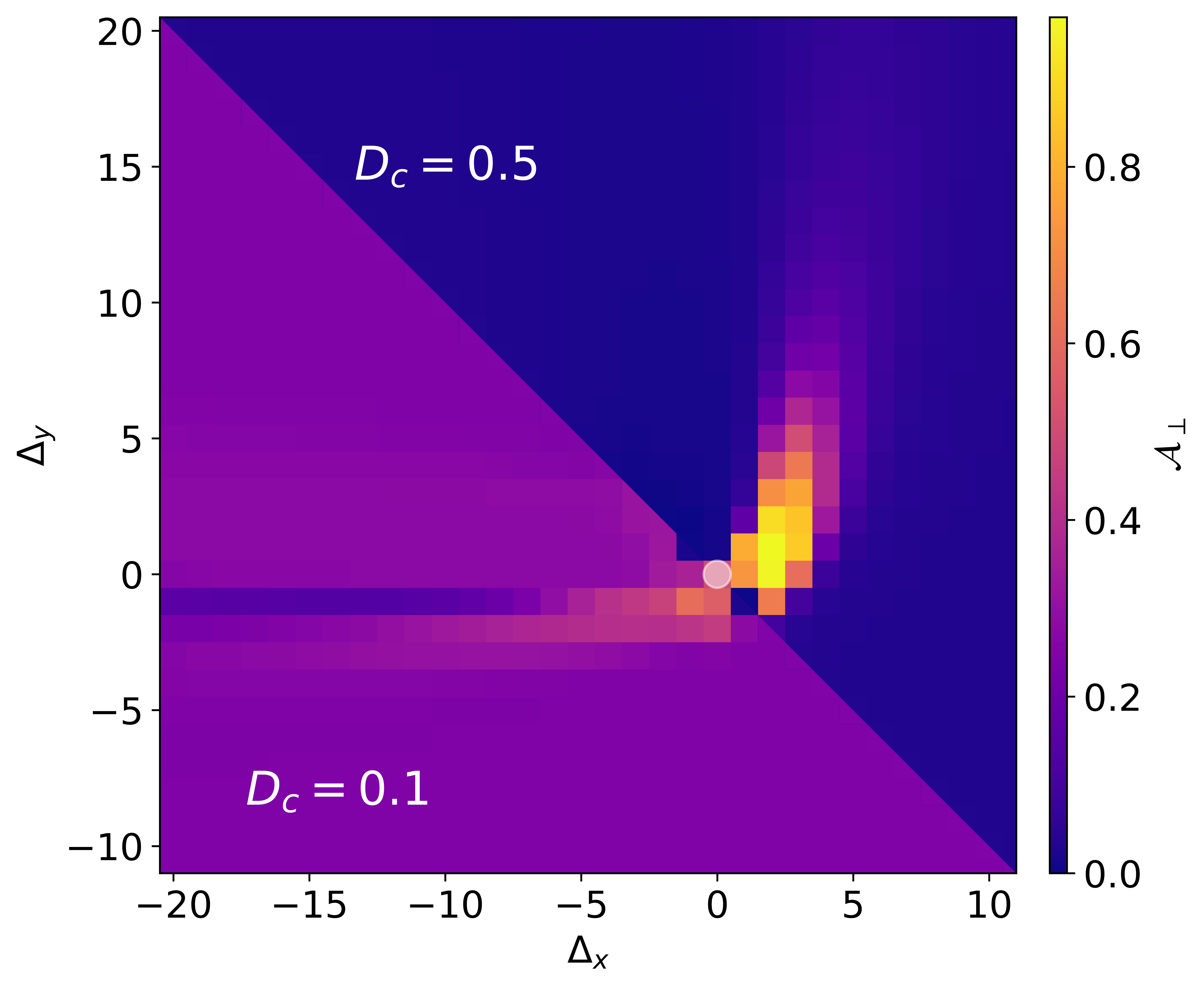}
	\end{center}
	\caption{Alignment probability $\mathcal{A}_\perp$ for $\beta = 100$. Because of the symmetry of the field by reflection with the line $\Delta_y = - \Delta_x$, we show it for $D_c = 0.1$ in the lower left plane and $D_c= 0.5$ in the upper right plane. The position $\mathbf{\Delta}=0$ is indicated by the white circle.}
	\label{fig:align_prob}
\end{figure}

First, let us tackle the case $\theta = \pi/2$ where the two walkers go in perpendicular directions. We compute the alignment probability $\mathcal{A}_\perp(\mathbf{\Delta})$, defined as the probability for two walkers at a distance $\mathbf{\Delta}$ to align their directions at the next step, given the the combined concentration field $c^{(2)}_{\infty}(\pi/2, \mathbf{\Delta})$. As shown in figure \ref{fig:align_prob}, the space dependence of the alignment probability field depends on the diffusion constant $D_c$ in a non-trivial way. First, consider the asymptotic value $\mathcal{A}_{\perp 0} = \lim_{|\mathbf{\Delta}|\to\infty} \mathcal{A}_\perp$. This is a signature of the self-interaction which we have discussed in a previous paragraph. In fact, walkers far away from each other can align {\it by chance} simply because their self-produced field allows them to pick the same direction. This is however less likely as the persistence length $l_p$ is large, yielding a low value for $\mathcal{A}_{\perp 0}$. 

The effect of the interaction can however be quantified via $\Delta \mathcal{A}_{\perp}(\mathbf{\Delta}) = \mathcal{A}_{\perp}(\mathbf{\Delta}) - \mathcal{A}_{\perp 0}$. For $D_c\to 0$, it takes non-negligible values along a long trail behind the walker but does not reach very large values. However for intermediate values such as $D_c=0.5$, $\Delta \mathcal{A}_{\perp}$ takes high values in a rather small region with maximal values slightly in front of the walker and to its right (or to its left for $\theta=-\pi/2$). Finally, as $D_c$ becomes very large, the cue diffuses so fast that the concentration gradient is too small and the alignment probability becomes homogeneous with values corresponding to the alignment probability of two independent blind random walks. 

\subsubsection{$\theta = \pi$}
\begin{figure}
	\begin{center}
		\includegraphics[width=\linewidth]{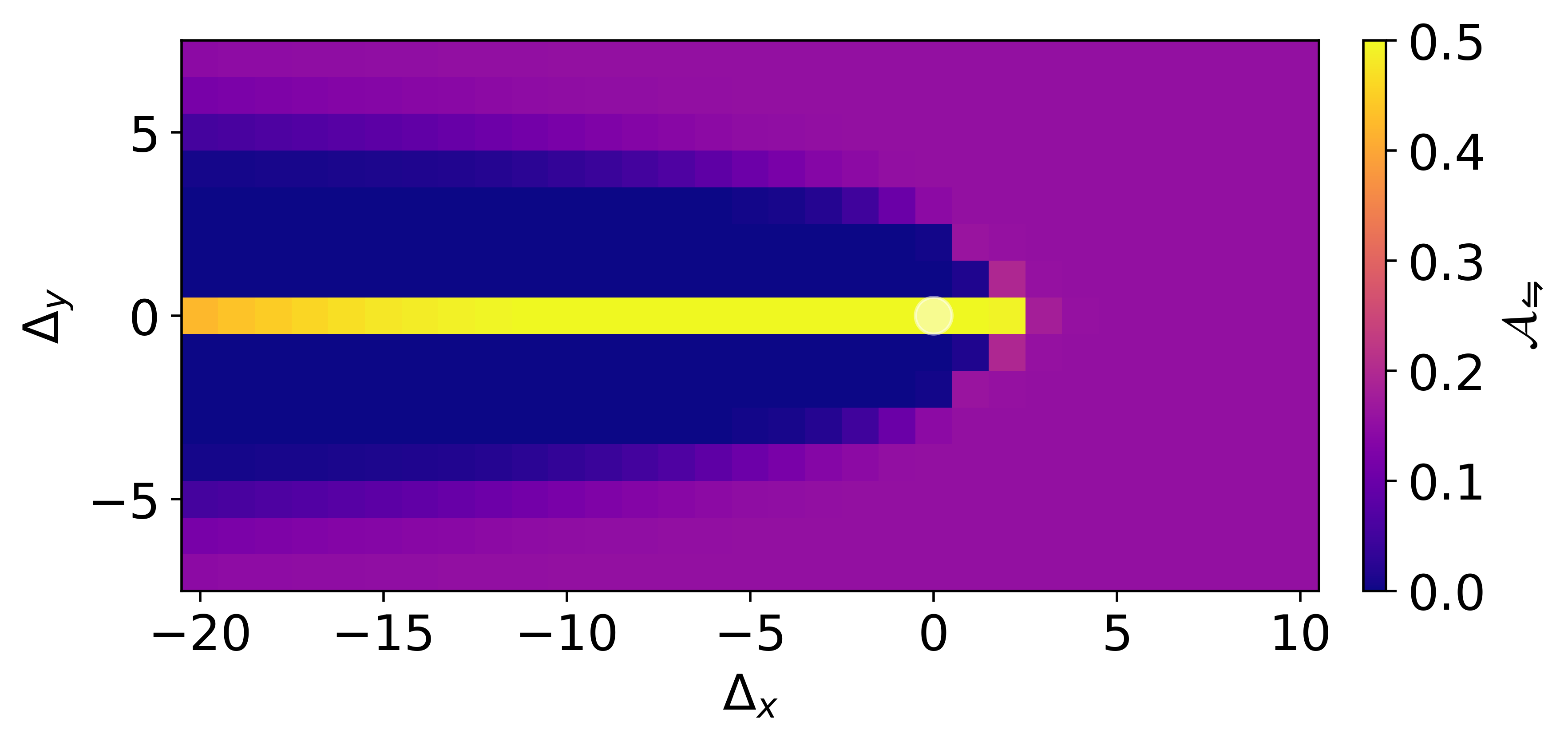}
		\includegraphics[width=\linewidth]{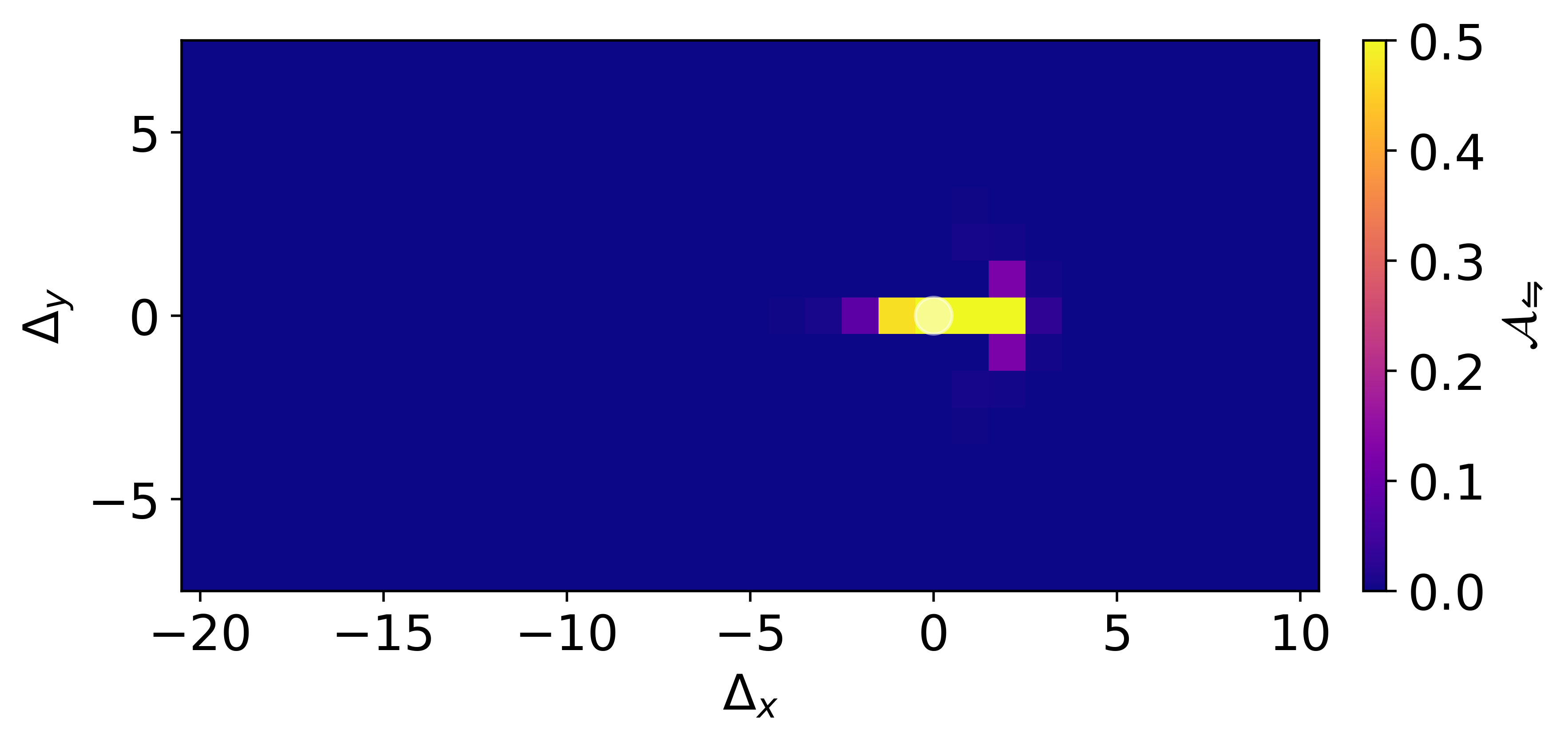}
	\end{center}
	\caption{Alignment probability $\mathcal{A}_\leftrightharpoons$ for $\beta = 100$ and $D_c=0.1$ (upper panel) and $D_c=0.5$ (lower panel). }
	\label{fig:align_prob_2}
\end{figure}

Next, let us consider the case $\theta=\pi$ where the two walkers go in parallel opposite directions. We can perform the same analysis as for $\theta = \pi/2$ and compute the probability $\mathcal{A}_\rightleftharpoons(\mathbf{\Delta})$ that two walkers at a mutual distance $\mathbf{\Delta}$ would align based on the combined field $c^{(2)}_\infty$. Due to the symmetry of the problem, the alignment interaction is weaker in this case, as there will be at most a probability $1/2$ for the two walkers to align. In particular, the alignment probability $\mathcal{A}_\leftrightharpoons$ is noticeably increased when the two walkers approach each other perfectly aligned, i.e. for $\Delta_y=0$ and $\Delta_x>0$, as shown in figure \ref{fig:align_prob_2}. However, $\mathcal{A}_\rightleftharpoons(\mathbf{\Delta})$ takes remarkably low values in a region corresponding to the back of the walker. This indicates that two walkers will probably align only if $\mathbf{\Delta}\cdot\mathbf{u}_{1,2} = 0$, but will otherwise scatter away. Note that this effect is much more pronounced for small values of $D_c$.

\subsection{Alignment stability}

Let us consider now the case $\theta=0$ where the two walkers move in the same direction and are separated by a distance $\mathbf{\Delta}$. As they are already aligned in this configuration, we want to quantify here the stability of the alignment. To formalize this, we define the persistence length $l_p^{(2)}(\mathbf{\Delta})$ as the mean number of steps the two walkers separated by a vector $\Delta$ will continue along the same direction before turning, given the initial stationary concentration field $c^{(2)}_{\infty}$. It is given by
\begin{equation}
l_p^{(2)}(\mathbf{\Delta}) = \sum_{k=0}^{\infty} k a_{\rightrightarrows}(\mathbf{\Delta})^k \left(1-a_{\rightrightarrows}(\mathbf{\Delta})\right) = \frac{a_{\rightrightarrows}(\mathbf{\Delta})}{1-a_{\rightrightarrows}(\mathbf{\Delta})}
\end{equation} 
where $a_{\rightrightarrows}(\mathbf{\Delta})$ is the probability for two walkers to continue in the same direction at the next time step:
\begin{equation}
a_{\rightrightarrows}(\mathbf{\Delta}) = z_{\rightrightarrows} e^{-\beta \left(c^{(2)}_{\infty}(0, \mathbf{\Delta}; \mathbf{e}_x) + c^{(2)}_{\infty}(0, \mathbf{\Delta}; \mathbf{\Delta}+\mathbf{e}_x)\right)}
\end{equation}
where we have used by convention $\mathbf{e}_x$ the common direction of the walkers and $z_{\rightrightarrows}$ is a normalization factor. 
Similarly to the cases $\theta \in \left\{ \pi/2, \pi\right\}$, this quantity might vary due to the self-interaction of a walker with its own field, but the value of $l_p$ corresponding to this effect is found for $|\mathbf{\Delta}|\to\infty$. 

\begin{figure}
	\begin{center}
		\includegraphics[width=\linewidth]{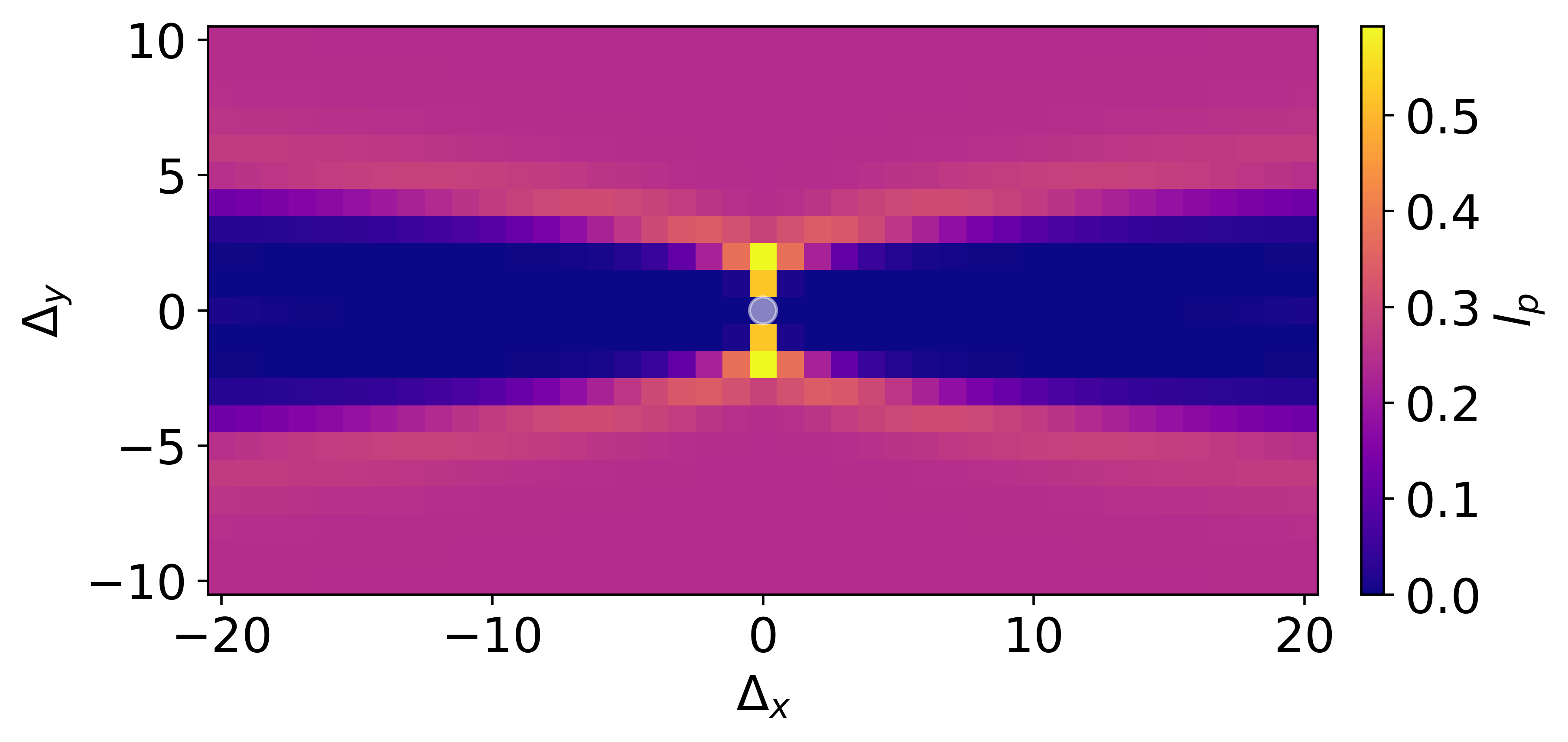}
		\includegraphics[width=\linewidth]{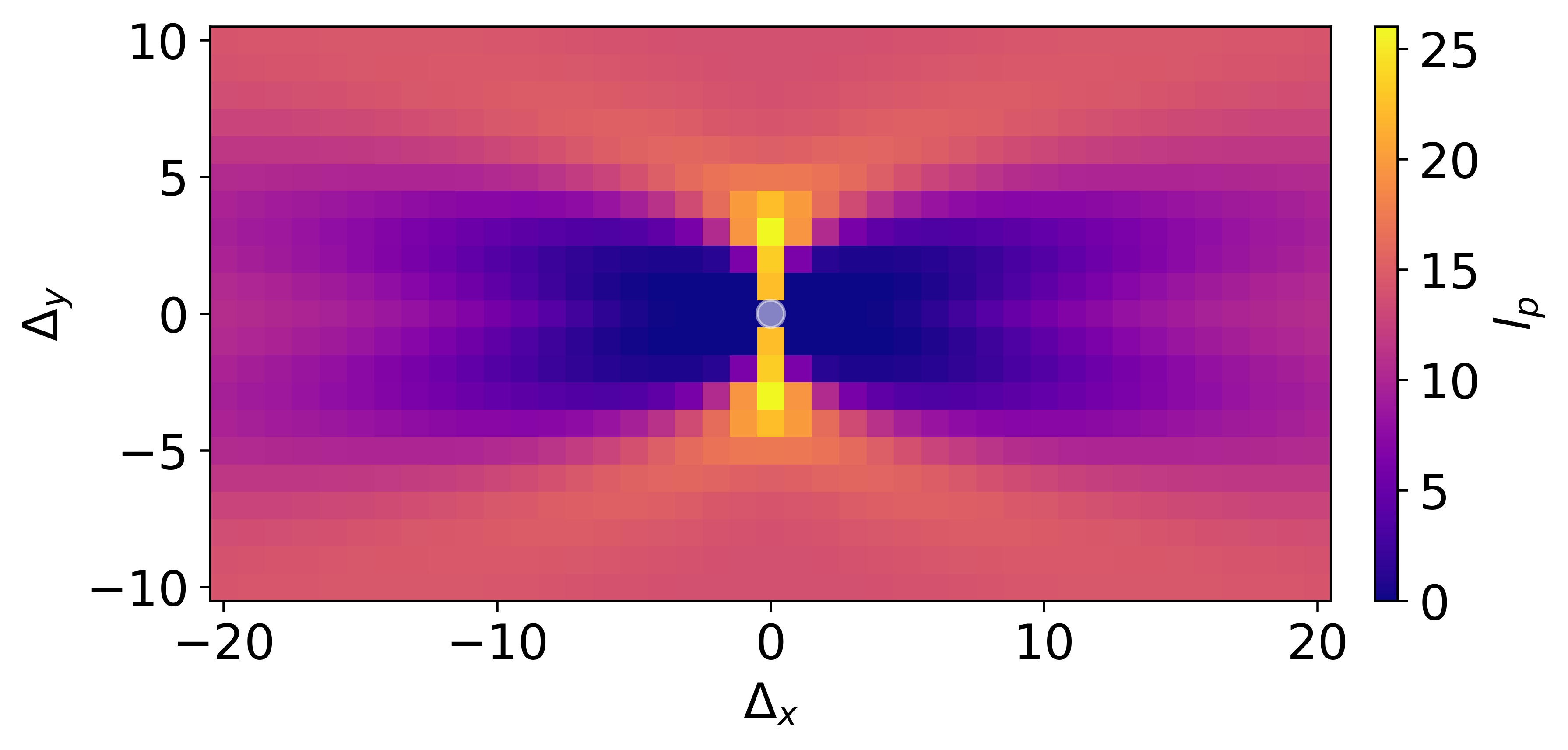}
	\end{center}
	\caption{Persistence length $l_p^{(2)}$ for $\beta = 100$ and $D_c=0.1$ (upper panel) and $D_c=0.5$ (lower panel). }
	\label{fig:persistence_length}
\end{figure}

\begin{figure}
\begin{center}
	\includegraphics[width=\linewidth]{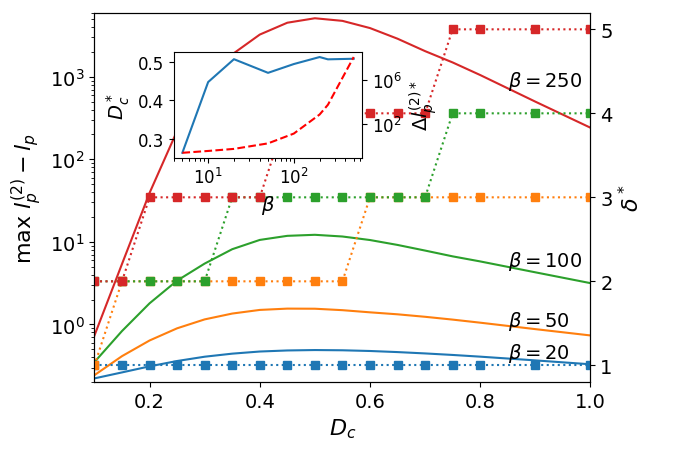}
\end{center}
\caption{$\max_\mathbf{\Delta} l_p^{(2)} - l_p$ (solid lines) and $\delta^*$ (dotted lines with squares) as a function of $D_c$ for various values of $\beta$. The inset shows the value of $D_c$ that maximizes this function (solid line) and the corresponding value of the persistence length from which is subtracted the persistence length due to self-interaction (dashed line).}
\label{fig:lp2max}
\end{figure}

We show in figure \ref{fig:persistence_length} two instances of $l_p^{(2)}(\mathbf{\Delta})$ for $\beta=100$, and $D_c=0.1$ and $D_c=0.5$. We note that there exists a maximum value of $l_p^{(2)}$ for $\mathbf{\Delta} = \pm \delta^* \mathbf{e}_y$ which depends on the parameters of the model but increases with $D_c$. Again, the value of the persistence length at this location strongly depends on $D_c$ and is maximized for a certain value of $D_c$. Similarly to the persistence length to due the self-interaction, this maximum $l_p^{(2)*}$ and its location $D_c^*$ depend on $\beta$ as we show in figure \ref{fig:lp2max}. 

\section{Results}

We performed Monte-Carlo simulations of the stochastic process defined in II on a 2D square lattice of size $L_x\times L_y$ and particle density $\varrho = N_w/L_xL_y$, with periodic boundary conditions. The walkers were initially placed one by one on randomly chosen empty sites. If $\varrho>1$, we first place $\left\lfloor \frac{N_w}{L_xL_y} \right\rfloor$ on each lattice site and then place the remaining walkers on random sites, provided that each site contains at most  $\left\lfloor \frac{N_w}{L_xL_y} \right\rfloor + 1$ walker. The concentration field was initially set to zero on all sites. The diffusion equation for the concentration field was solved using the Crank-Nicolson method with alternating direction, using an integration timestep $\delta t = 0.01 D_c^{-1}$. 
	
\subsection{Dilute regime}

\begin{figure}
	\begin{center}
		\includegraphics[width=0.99\linewidth]{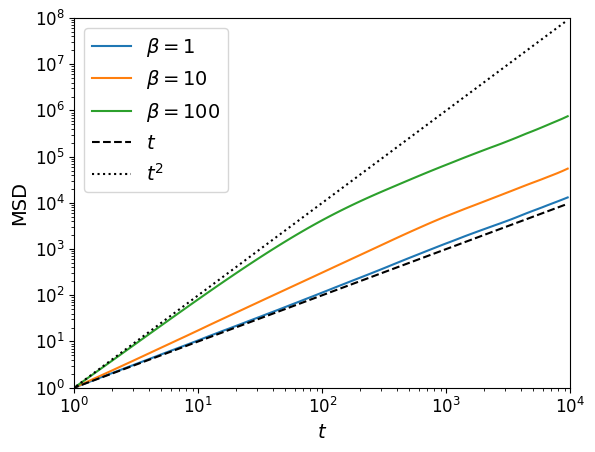}
		\caption{Mean-square displacement of a single walker for $D_c=1$ and various values of $\beta$. The limit cases $\left\langle \mathbf{r}(t)^2 \right\rangle =t$ and $\left\langle \mathbf{r}(t)^2 \right\rangle =t^2$ are shown in dashed and dotted lines, respectively.}
		\label{fig:MSD_single}
	\end{center}
\end{figure}

We start by investigating the low-density regime where we place only one walker in the simulation box, leading to a number density of $\varrho = 1/L_xL_y$, in order to characterize the effect of the interaction of a walker with its own concentration field on its dynamics. To this end, we compute the mean-square displacement (MSD) of the process, that we show in figure \ref{fig:MSD_single}. 

From the analysis presented in the previous section, we expect the field to act as an aligning force for the walker. This implies that for $\beta \to \infty$ and $D_c>0$, we expect the walk to be ballistic, such that $\left\langle \mathbf{r}(t)^2 \right\rangle = t^2$, which is equivalent to an infinite persistence length. The limit $\beta \to\infty$ and $D_c = 0$ must however be taken with care. Here, the ballistic regime can not be reached as the chemical does not diffuse. As the walker proceeds, there is on average no difference between the forward, left and right sites such that the walker can easily take turns and break the ballistic regime. On the other extreme, for $\beta=0$, the process is a conventional random walk such that the MSD is entirely diffusive, i.e.  $\left\langle \mathbf{r}(t)^2 \right\rangle = t$. 

For finite values of $\beta$, we observe an intermediate behavior. The MSD is super-diffusive at short times reaches a diffusive regime at long times with $\left\langle \mathbf{r}(t)^2 \right\rangle = D_w t$. We show an example of the MSD for $D_c = 0.1$ and various values of $\beta$ in figure \ref{fig:MSD_single}. Note that even for the single-walker case, the numerical computations of the MSD suffers from finite-size effects as the walker can potentially interact with the concentration field produced by one of its periodic images. However, if the simulation box is large enough this field must have sufficiently diffused away such that the total field that the walker feels is negligibly impacted by the field produced by periodic images. 

\subsection{Dense regime}

We consider now arbitrarily large number densities. The phenomenological analysis discussed previously indicate that walkers can effectively interact over length scales that depend on $D_c$ in a non-trivial way. While an alignment interaction at short distance can order the system, a similar interaction over long distances can however prevent large ordered regions from spontaneously emerging. 

\begin{figure}
	\begin{center}
		\includegraphics[width=0.99\linewidth]{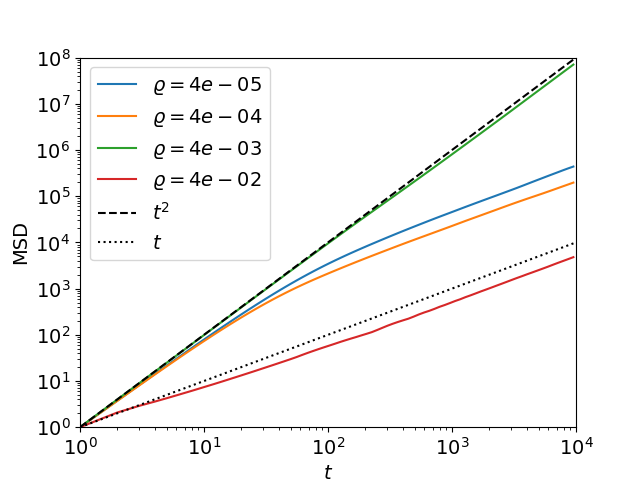}
		\caption{Mean-square displacement for $D_c=1$, $\beta=100$ and various values of $\varrho$. The limit cases $\left\langle \mathbf{r}(t)^2 \right\rangle =t$ and $\left\langle \mathbf{r}(t)^2 \right\rangle =t^2$ are shown in dashed and dotted lines, respectively.}
		\label{fig:MSD_dense}
	\end{center}
\end{figure}

First, we compute again the MSD for increasing numbers of walkers, as shown in figure (\ref{fig:MSD_dense}) for $D_c=1$,  $\beta=100$, $L_y=100$ and $L_x=400$. As the density is low, the behavior observed in the single-walker case still holds, namely an initial super-diffusive behavior followed by a diffusive one. The transition time between the two regimes is however decreased as density increases since the super-diffusive behavior can persist on a timescale related to the average volume per particle, while it is broken because of spontaneous fluctuations in the single-walker case.

As the density increases even more, the MSD exhibits a non-trivial behavior. In the specific case displayed in figure (\ref{fig:MSD_dense}), some values of $\varrho$ result in a persistent super-diffusive regime with an exponent very close to $2$. This indicates a clear change in the motile behavior of the walkers and is the signature of a new phase of the system. For even larger densities, this phenomenon disappears and we observe an almost purely diffusive behavior, with even a sub-diffusive regime at intermediate times.  

To assess the existence of different phases for intermediate values of $\varrho$ and characterize the symmetry breaking of the system, we define a global order parameter $\sigma$ as
\begin{equation}
\sigma = \frac{N_\uparrow+N_\downarrow}{N_\rightarrow + N_\leftarrow}
\end{equation}
where $N_{\uparrow, \downarrow, \leftarrow, \rightarrow}$ refers to the number of walkers traveling in one the four directions, provided that $N_\uparrow+N_\downarrow<N_\leftarrow+N_\rightarrow$ such that $0\leq\sigma\leq 1$. This quantity is devised to indicate whether the four-fold symmetry is broken. To detect indications of pattern formation, we also compute the discrete Fourier transforms $\hat{\rho}_\mathbf{v}(k_x, k_y)$ of all 4 directed density fields $\rho_\mathbf{v}(x,y)$, defined as 
\begin{equation}
	\hat{\rho}_\mathbf{v}(k_x, k_y) = \sum_{x = 0}^{L_x-1} \sum_{y = 0}^{L_y-1} e^{- i2\pi (\frac{k_x x}{L_x}  + \frac{k_y y}{L_y})} \rho_\mathbf{v}(x,y)
\end{equation}
Note that $\sigma = \frac{ |\hat{\rho}_{\mathbf{e}_y}(0, 0)|+|\hat{\rho}_{-\mathbf{e}_y}(0, 0)|}{|\hat{\rho}_{\mathbf{e}_x}(0, 0)|+|\hat{\rho}_{-\mathbf{e}_x}(0, 0)|}$, provided that there are more walkers going along the $x$-axis than along the $y$-axis. We also introduce $\psi_\mathbf{v}(q)$ defined as $\psi_\mathbf{v}(q) = |\hat{\rho}_\mathbf{v}(q/L_x, 0)|$ if $\mathbf{v}=\pm \mathbf{e}_x$ and $\psi_\mathbf{v}(q) = |\hat{\rho}_\mathbf{v}(0, q/L_y)|$ if $\mathbf{v}=\pm \mathbf{e}_y$. We use $\sigma$ and $\rho_\mathbf{v}$ to identify different phases in the $(\rho,\beta)$ plane, which we show in Fig. 9 and describe in detail now.
\begin{figure*}
	\begin{center}
		\begin{subfigure}{.32\linewidth}
			\includegraphics[width=\linewidth]{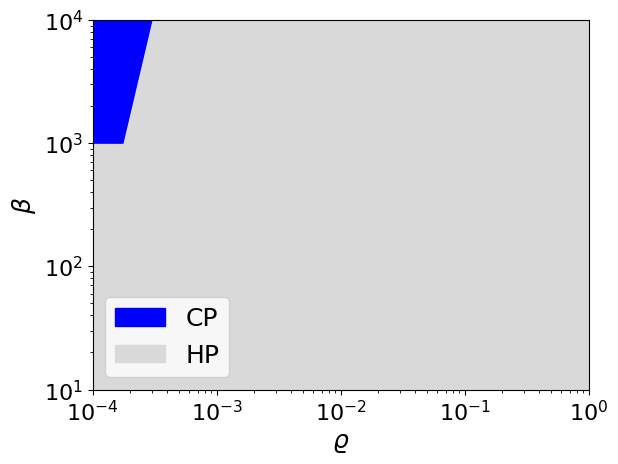}
			\caption{$D_c=0.1$}
			\label{fig:diag_Dc01}
		\end{subfigure}
		\begin{subfigure}{.32\linewidth}
			\includegraphics[width=\linewidth]{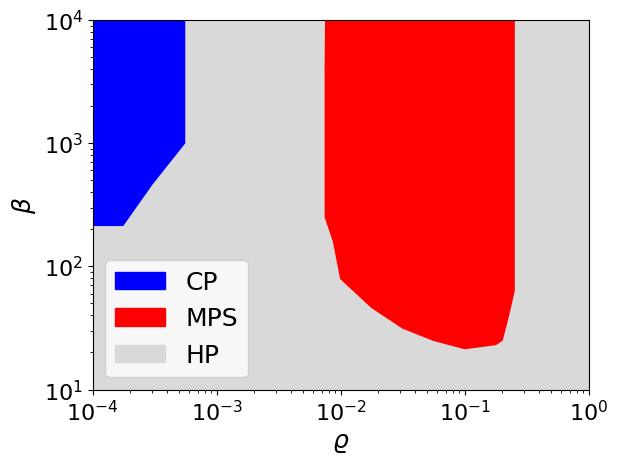}
			\caption{$D_c=0.5$}
			\label{fig:diag_Dc05}
		\end{subfigure}
		\begin{subfigure}{.32\linewidth}
			\includegraphics[width=\linewidth]{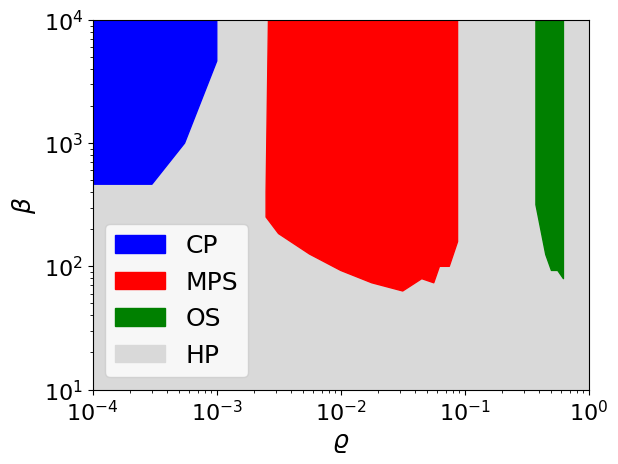}
			\caption{$D_c=1$}
			\label{fig:diag_Dc1}
		\end{subfigure}
		\caption{Diagrams in the $(\varrho, \beta)$ plane showing various phases found in the system for three different values of the diffusion constant $D_c$. Note that both axes are in logarithmic scale.}
		\label{fig:phase_diagram}
	\end{center}
\end{figure*}

\subsubsection{Homogeneous phase (HP)}

We refer to any system with $\bar{\sigma}>0.5$ as the homogeneous phase. Here, no patterns are formed and no symmetry is broken on large length scales. We will not discuss this "gas" phase in more details in this paper and focus on the phases for which we observe broken symmetries. 

\subsubsection{Cluster phase (CP)}

\begin{figure}
	\begin{center}
		\includegraphics[width=\linewidth]{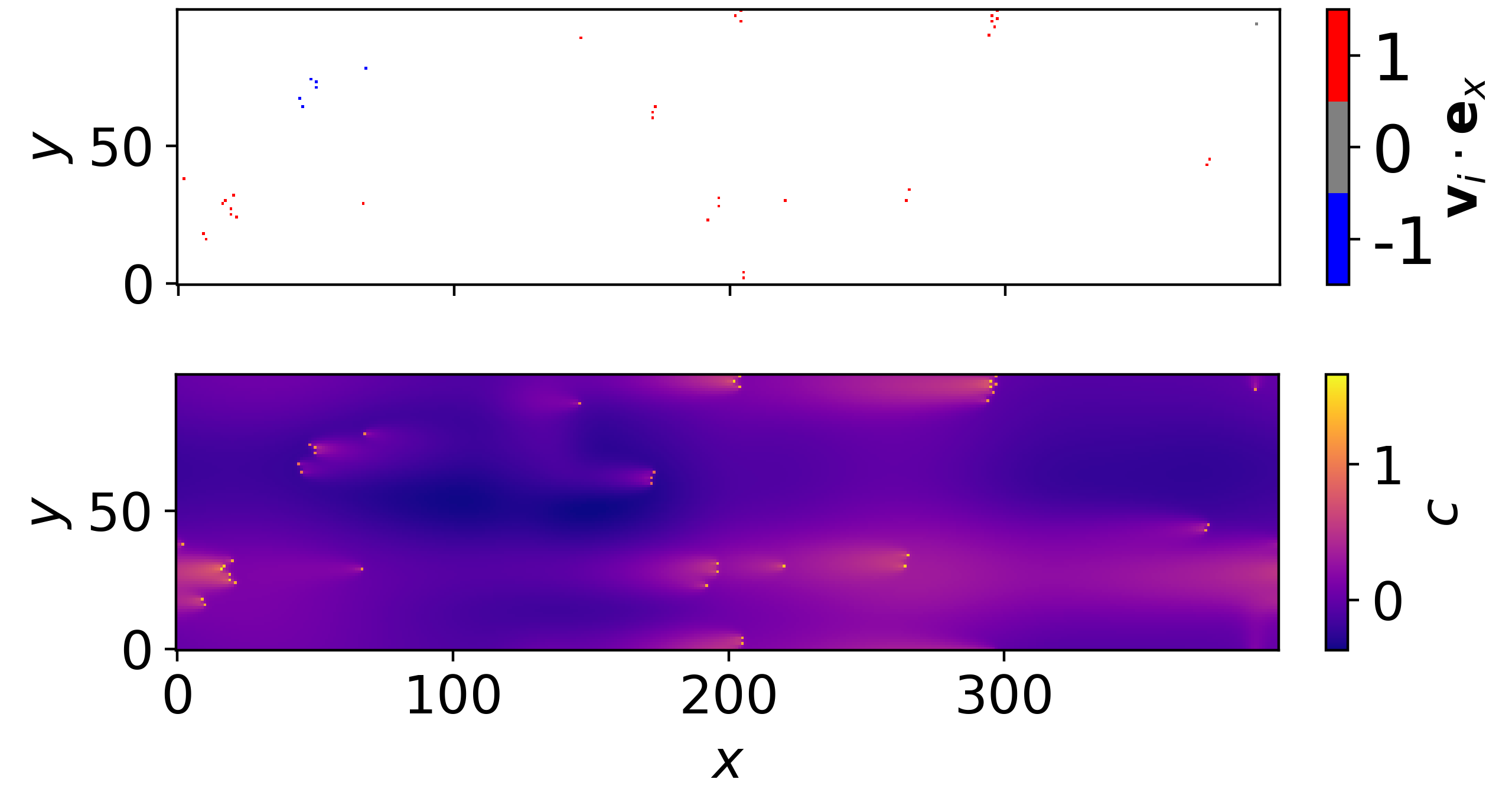}
		\caption{Typical configuration of the cluster phase ($\varrho=0.001$, $\beta=4640$, $D_c=0.5$, $L_x=400$, $L_y=100$). The upper panel shows the positions of walkers and their traveling directions along the $x$-axis while the lower panel shows the corresponding concentration field.}
		\label{fig:cluster_phase}
	\end{center}
\end{figure}

For very low densities and large values of $\beta$, we find a phase where most particles travel along the same axis, in either direction. We show in figure \ref{fig:cluster_phase} a typical snapshot of such a phase. Here, particles group in small clusters which behave essentially as ballistic units. We identify this phase as the mean value of $\sigma$ is lower than $0.5$ but no particular peak is observed in $|\hat{\rho}_\mathbf{v}(k_x, k_y)|$. To explain the formation of the clusters and their stability, we first recall that without interaction between walkers, large values of $\beta$ lead to ballistic motion. When multiple walkers are placed together, the ballistic motion can be broken if two walkers traveling in different directions can sense each other's concentration field. Here, the formation of the cluster phase can be summarized as follows : 

Initially, particles walk ballistically as long as they do not sense the field of another walker. This can happen over rather large distances due to the overall low density. Eventually, two walkers going in orthogonal directions will reach a mutual distance such that the alignment interaction will be high and the two walkers will adopt the same direction. Because of the large values of $\beta$, these two walkers should not come apart and will therefore form a cluster of 2 particles as long as they do not encounter a third particle or an other cluster. In such a case, either they will absorb the third particle or merge with the other cluster, or be split in parts. The formation and growth of such clusters will however stop whenever all clusters (or single particles left alone) travel along the same axis and are located at large enough distances. Then, they will never interact again and continue along their ballistic paths indefinitely. This is however possible only because the overall density is very low such that a configuration can be spontaneously reached.

In accordance with the phenomenological analysis made earlier, we note that the minimal value of $\beta$ from which the dilute ballistic phase can be formed depends non-monotonically with $D_c$, as the persistence length and hence the strength of the ballistics behavior is maximized for $D_c \simeq 0.5$ for $\beta \gg 1$.

\subsubsection{Macro phase separation (MPS)}

At intermediate densities, we observe a second phase where system-spanning bands are formed. Particles arrange densely in a narrow region and travel ballistically in a direction perpendicular to the band which connects to its periodic image, as shown in figure \ref{fig:band_instances}. For bands traveling along the $x$-axis, the mean value of $\sigma$ is much lower than 0.5, and $|\hat{\rho}_\mathbf{v}(k_x, 0)|$ for decays over an inverse length scale corresponding to the band size. For dense rectangular bands, it also presents peaks similar to a sinc function. This set of features allows to formally identify the phase.

In this region of the phase diagram, the system can reach three different stationary states :
\begin{enumerate}[label=\roman*]
	\item One single band is formed and all particles travel along the same direction.
	\item One single band is formed but a substantial amount of particles still travel in the opposite direction, in a dilute cloud. 
	\item Two bands of the same size travel in opposite directions.
\end{enumerate}
A single set of parameters $\left(\varrho, D_c, \beta\right)$ can spontaneously lead to these three configurations, with different probabilities. We show in figure \ref{fig:band_instances} typical timeseries of $N_\rightarrow$, $N_\uparrow$, $N_\leftarrow$ and $N_\downarrow$ in the three scenarios. In cases (ii) and (iii), the spikes occurring at periodic intervals correspond to collisions between the two structures traveling in opposite directions. The phenomenology of these events can be summarized as follows:
\begin{figure*}
	\begin{center}
		\begin{subfigure}[b]{\linewidth}
			\centering
			\includegraphics[width=.95\columnwidth]{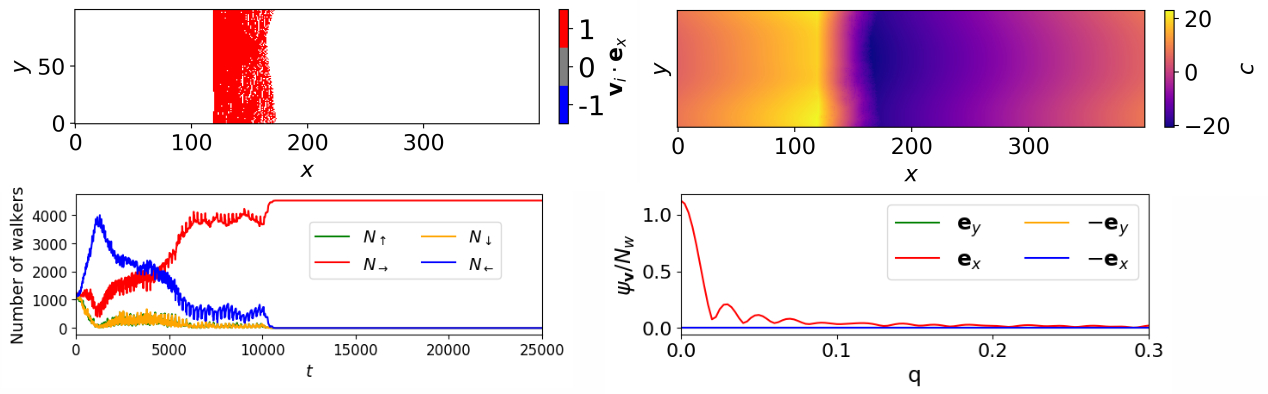}
			\caption{$\varrho=0.11$, $\beta=1000$, $D_c=0.5$. One single band has formed.}
			\label{fig:oneband}
		\end{subfigure}
		\begin{subfigure}[b]{\linewidth}
			\centering
			\includegraphics[width=.95\columnwidth]{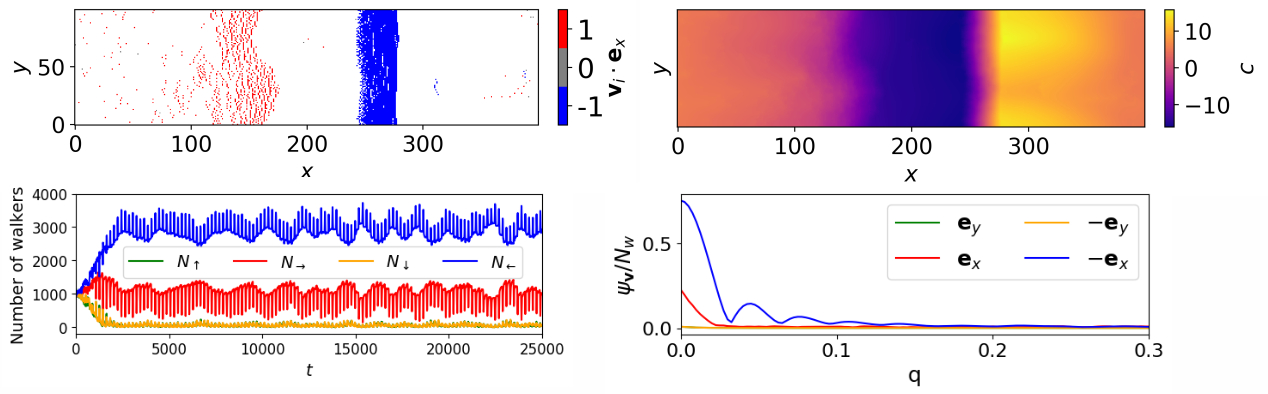}
			\caption{$\varrho=0.1$, $\beta=215$, $D_c=0.5$. One band is traveling to the left while a more dilute cloud is traveling to the right.}
			\label{fig:onetwo_bands}
		\end{subfigure}
		\begin{subfigure}[b]{\linewidth}
			\centering
			\includegraphics[width=.95\columnwidth]{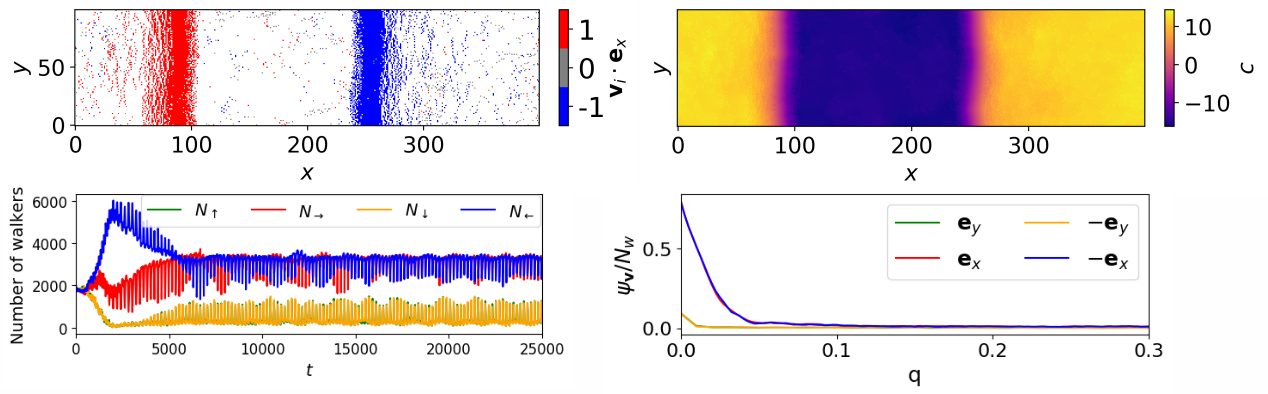}
			\caption{$\varrho=0.18$, $\beta=100$, $D_c=0.5$. Two bands of similar profiles are traveling in opposite directions. }
			\label{fig:two_bands}
		\end{subfigure}		
		\caption{Different formations of bands. In each example we show snapshots of the total orientation along the $x$-axis (upper left panel) and concentration field (upper right panel), with the corresponding timeseries of $N_\rightarrow$, $N_\uparrow$, $N_\leftarrow$ and $N_\downarrow$ as a function of time(lower left panel)and the profiles $\psi_\mathbf{v}$ for all 4 directions (lower right panel).}
		\label{fig:band_instances}
	\end{center}
\end{figure*}

\begin{figure}
	\begin{center}
		\includegraphics[width=.49\linewidth]{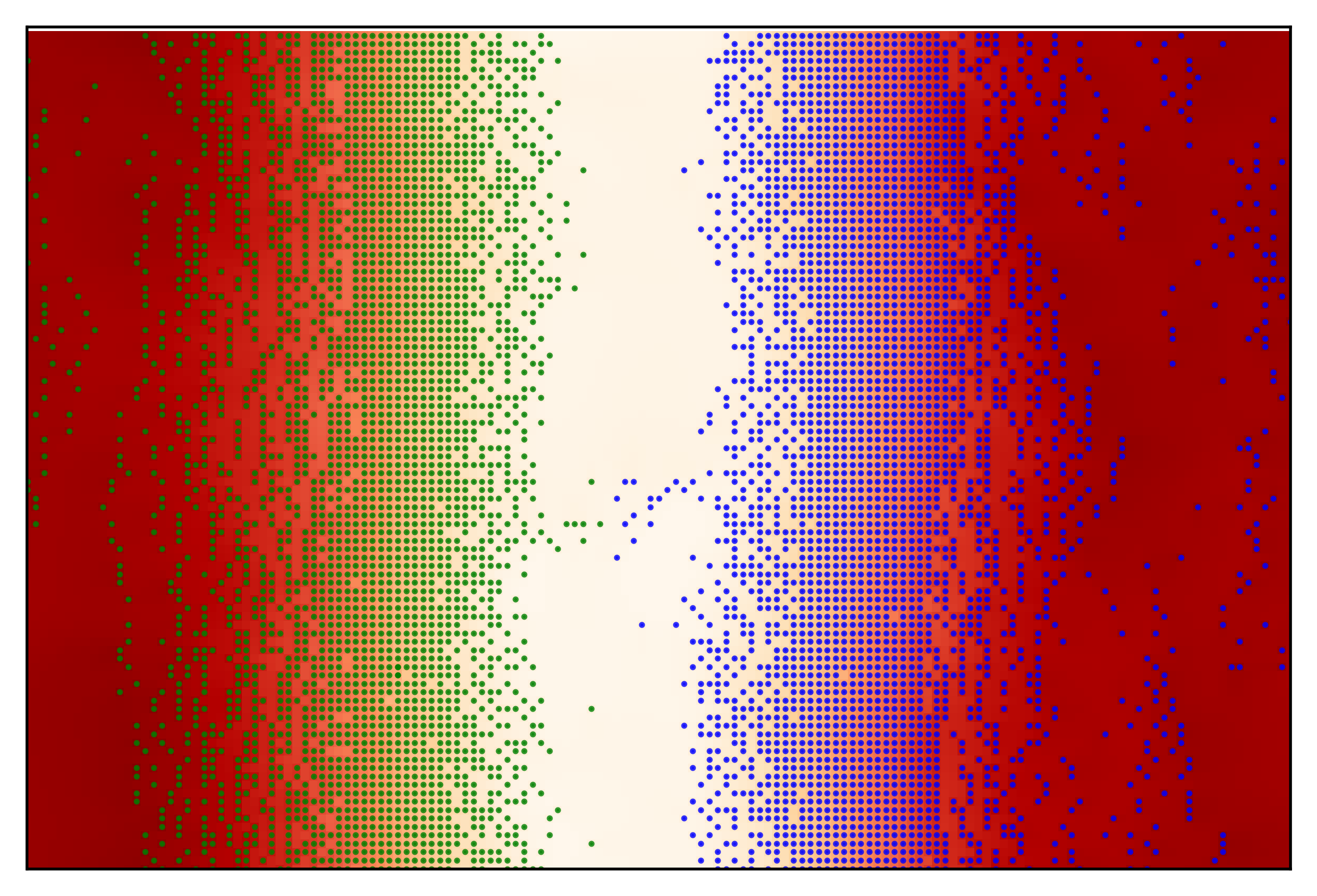}
		\includegraphics[width=.49\linewidth]{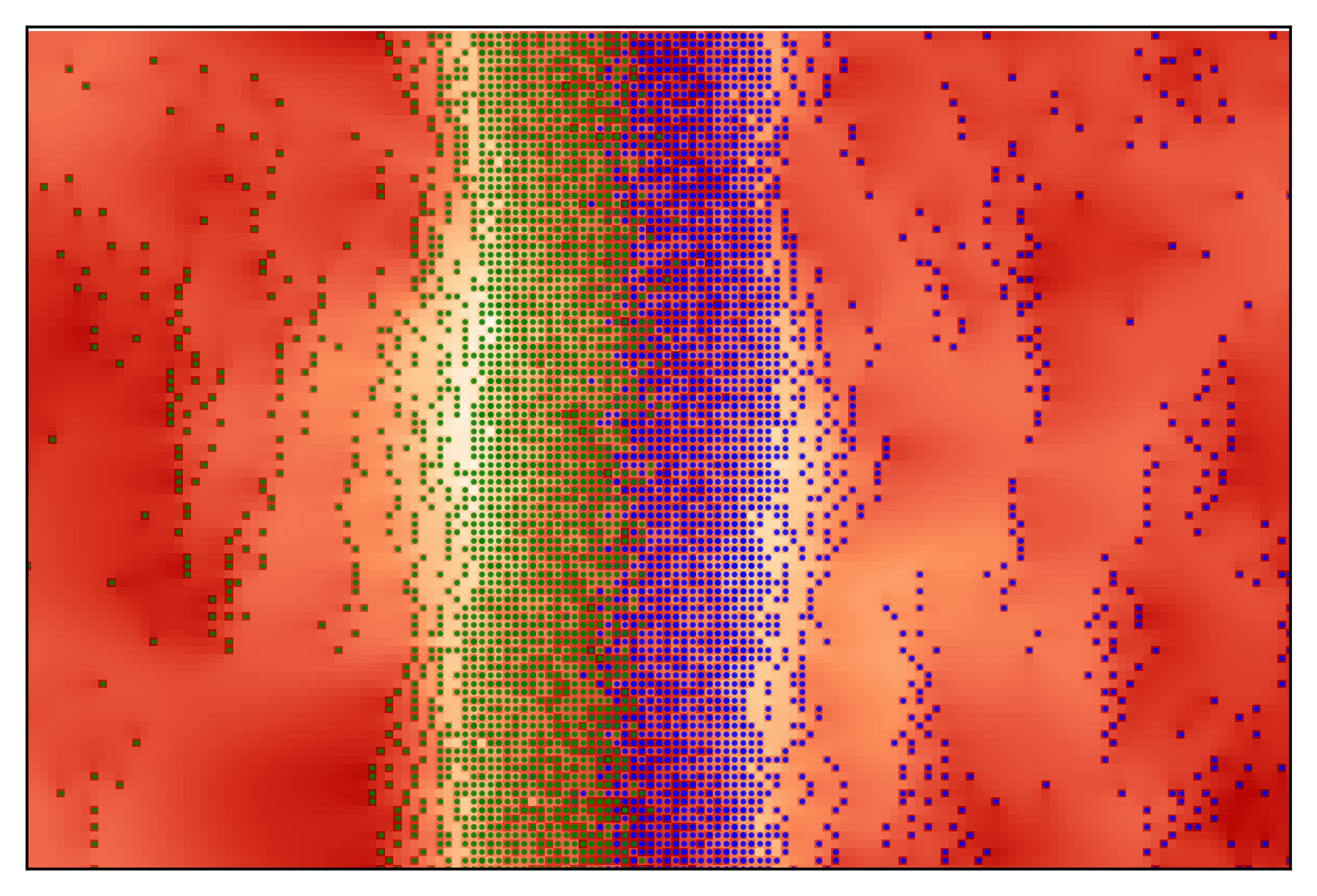}
		\includegraphics[width=.49\linewidth]{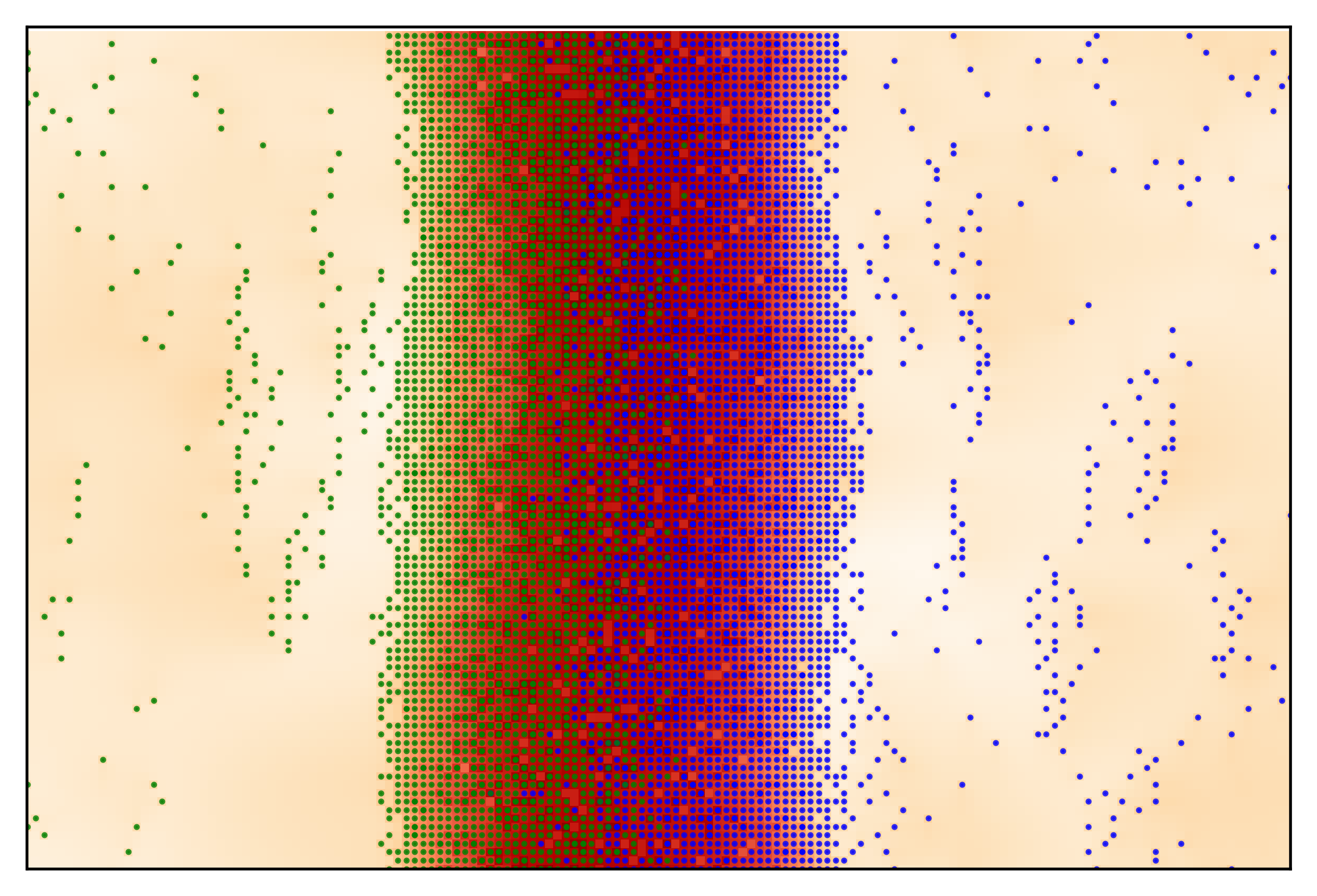}
		\includegraphics[width=.49\linewidth]{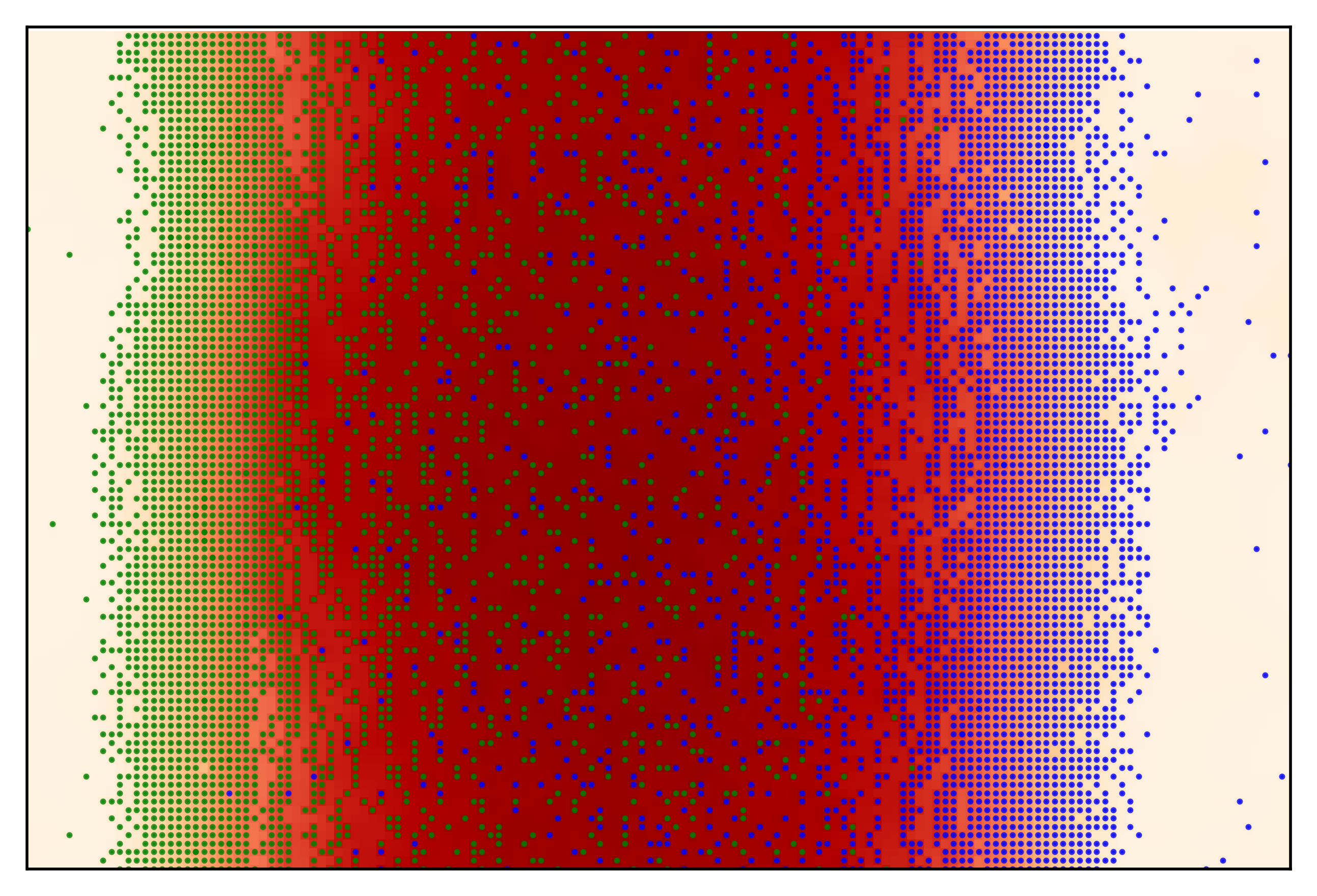}
		\caption{Series of snapshots in a bouncing event between 2 bands. The background color codes the value of the concentration field (note that the scale is not the same in all snapshots to increase the contrast), while the dots indicate the position of walkers, the color of which refers to their position in the first snapshot.}
		\label{fig:bounce_bands}
	\end{center}
\end{figure}

In the case (iii) of two colliding bands, the concentration field is more intense at the back of each bands as long as they are far apart, which pushes them independently in opposite directions. When the fronts of both bands collide, they temporarily merge to form a larger structure, the center of which becomes denser and more concentrated. As long as the concentration field on both sides of this super-band is more intense than at the center, the two bands will keep moving in their original direction, and the concentration will keep increasing at the center. Eventually, the overall concentration gradient will flip its direction such that the walkers will favorably reverse their orientations. Doing so, the super-band will split again and the two bands will reform as two independent structures going in opposite directions. Overall, the process can be understood as two bands bouncing on each other. We show in figure \ref{fig:bounce_bands} a series of snapshots showing this phenomenon. 

For the case (ii) of a collision between a band and a dilute cloud, the mechanism is essentially the same. When the front of both structures meet, the particle density and concentration field increase locally, where the collision occurs. However, since the concentration in the cloud is much lower than in the band, walkers in the cloud will quickly reverse their direction and be absorbed at the front of the band. On the other hand, the same increase in concentration will in turn lead some particles within the band to reverse their direction, hence propagating the local concentration increase in the direction opposite to the band velocity. When this wave reaches the back of the band, walkers in this region will eventually reverse their directions and leave the band to reform the dilute cloud. 

As seen in figure \ref{fig:phase_diagram}, macro phase separation occurs at intermediate values of the particle density and high values of $\beta$. In fact, given $\delta^*$ the typical length scale that maximizes the alignment stability between walkers, a system-spanning band can only form if $N_w \delta^*$ si larger than the smallest dimension of the simulation box. As shown in section III, $\delta^*$ increases with $D_c$.As a result, the lowest particle density $\varrho$ for which macro phase separation can be observed decreases with $D_c$. In addition, we have also shown that the alignment interaction is weaker as $D_c$ is larger than an {\it optimal} value $D_c^*$. A stronger coupling strength $\beta$ is therefore needed for the bands to sustain. This is illustrated in figure \ref{fig:phase_diagram} where the region of macro phase separation is shifted towards lower densities and larger values of $\beta$ as $D_c$ increases. 

We also remarkably notice that the macro phase separation is not observed as $D_c$ is very small. This is due to the very persistent trail left by the walkers which prevents any band-like structure to spontaneously form.

\subsubsection{Oscillating stripes (OS)}

\begin{figure}
	\begin{center}
		\includegraphics[width=\linewidth]{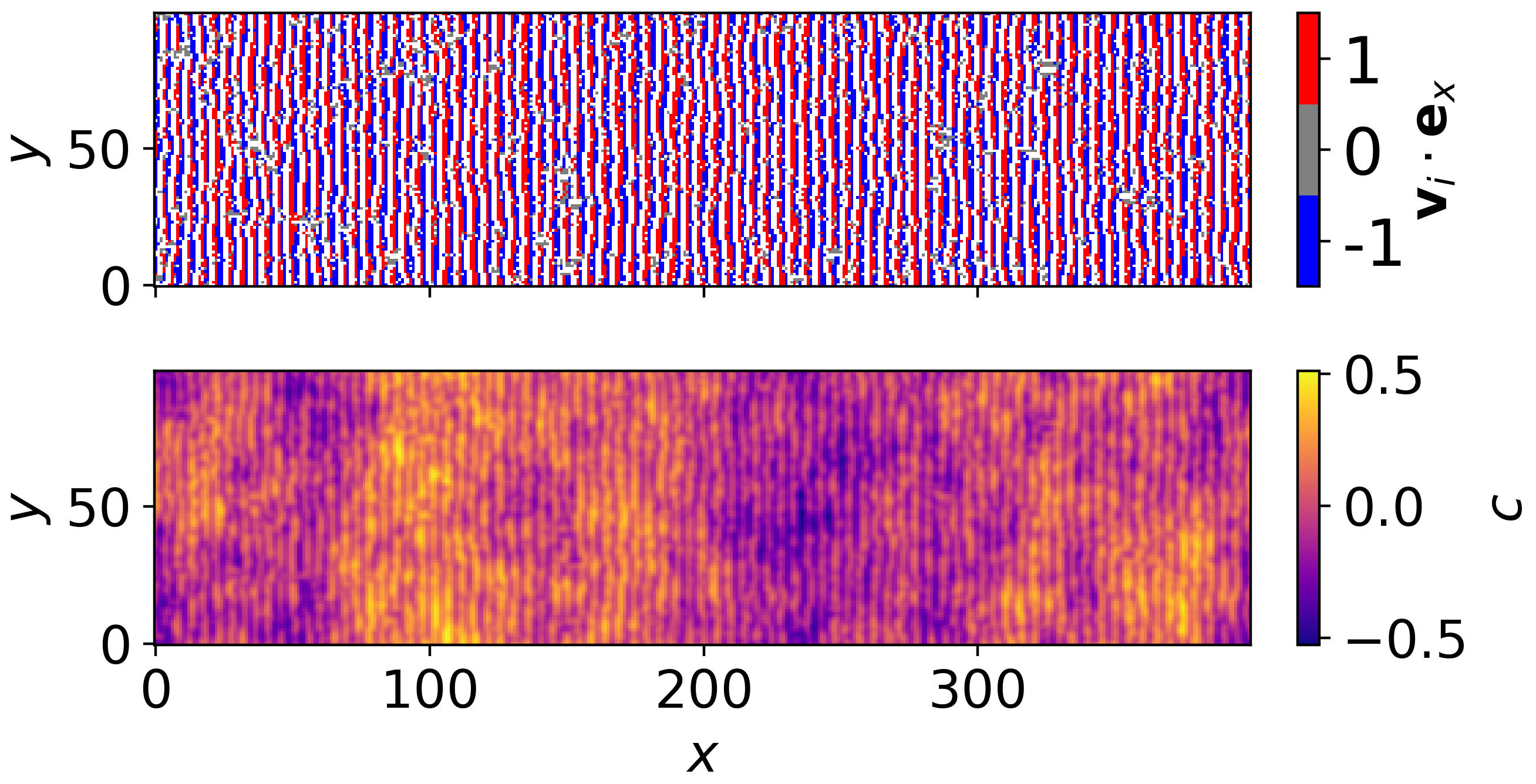}
		\includegraphics[width=\linewidth]{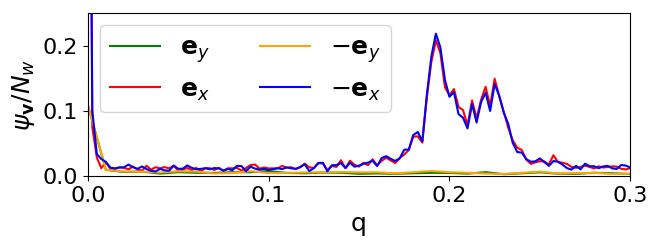}
		\caption{Snapshot of a system in the oscillating band phase ($\varrho=0.56$, $\beta=1000$, $D_c=1$). The upper panel shows the direction of particles along the $x$-axis while the lower panel shows the corresponding concentration field. We also show $\psi_\mathbf{v}$ for all 4 directions.}
		\label{fig:beating_phase}
	\end{center}
\end{figure}

In a narrow region of the phase diagram, and only for certain values of the diffusion constant, we observe a third ordered phase. It is composed of narrow stripes, typically 2-lattice-site wide, the left side of which travels to the left and the right side to the right. We show a typical configuration and the corresponding concentration field in figure \ref{fig:beating_phase}. At each time step, each band splits in two parts which recombine with half of the neighboring bands. This forms an oscillating "beating" pattern that is stable once it has been reached. We call this phase the {\it oscillating stripes} (OS). It can be clearly identified using again $|\hat{\rho}_\mathbf{v}(k_x, 0)|$ as peaks are observed for values of $k_x$ corresponding to the width of stripes and the spacing between them. This phase can also be seen as a stationary wave of the particle density field, as 2-walker-wide stripes travel at constant speed, namely two sites per timestep.

We note that this phase is only observed for sufficiently large values of the diffusion constant $D_c$. To understand this, we perform a stability analysis via the following experiment. First, we initialize the system by placing walkers in a perfect arrangement of 2-site-wide stripes separated by gaps of the same size. Then, we impose the oscillatory motion observed in the simulations by making the left side of a stripe move to the left and its right side move to the right, hence recombining neighboring half-stripes into new stripes. This is performed until the concentration field reaches a stationary profile. From this state, we pick one random walker and make it move to a neighboring site in an orthogonal direction and let the concentration field diffuse over one time step, creating a defect in the pattern. Then, we let the system evolve according to the actual rules of the model. 

For $\beta\to\infty$, we observe that the concentration field is such that the stripes reform immediately after the first time step if $D_c$ is larger than a certain threshold $D_c^*$. Otherwise, it will need more steps to reform or even destroy completely the overall structure. To estimate the value of $D_c^*$, we investigate the details of the concentration field after the creation of the defect, right before the system is let free to evolve. 

\begin{figure}
	\begin{center}
		\includegraphics[width=\linewidth]{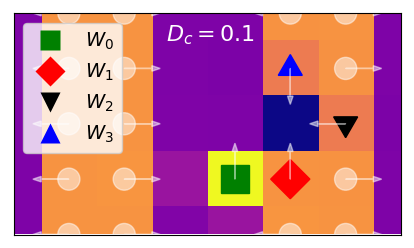}
		\includegraphics[width=.49\linewidth]{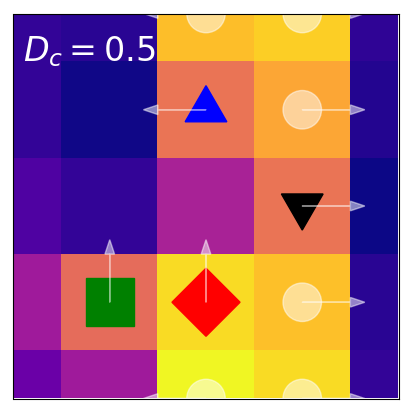}
		\includegraphics[width=.49\linewidth]{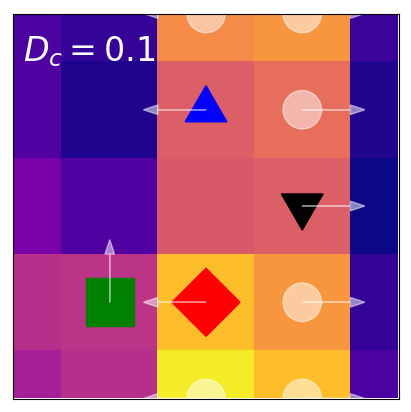}
		\includegraphics[width=\linewidth]{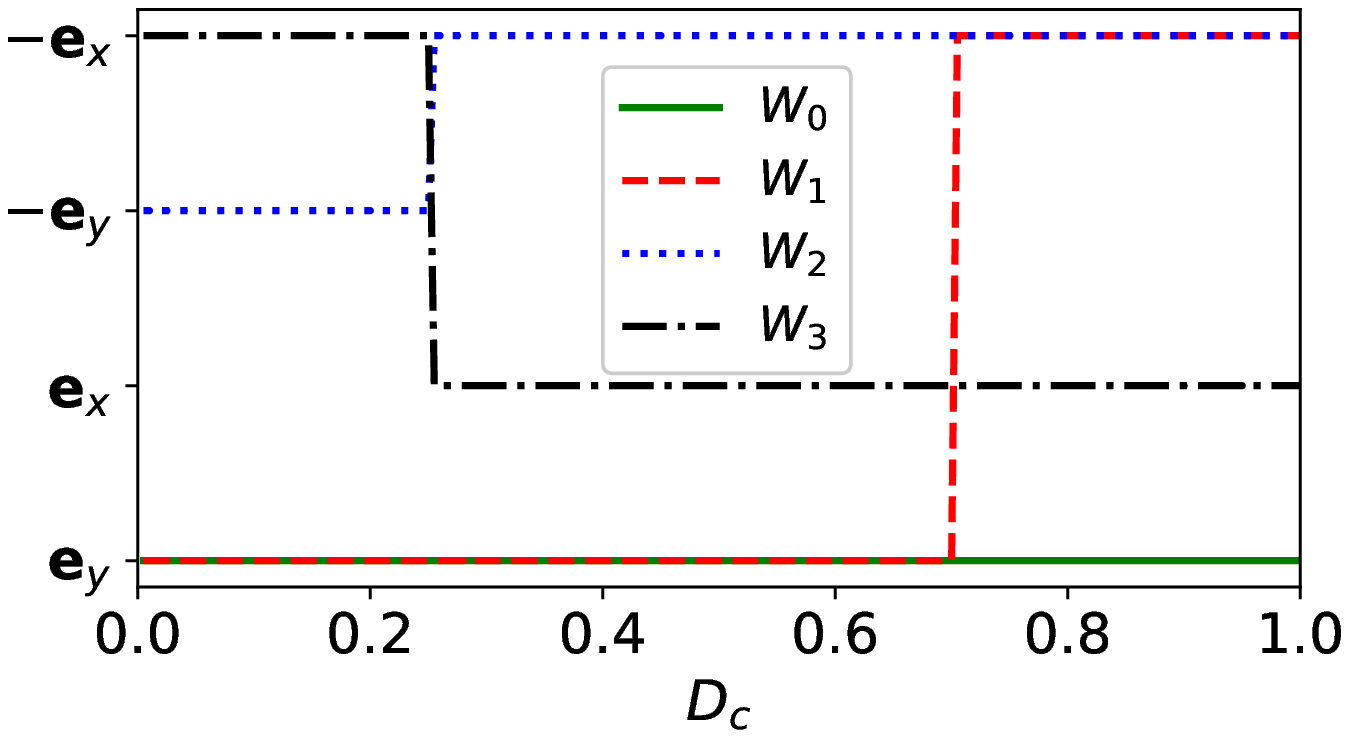}
		\caption{Setup for the stability analysis for three different values of $D_c$. The color codes for the concentration field. Positions of all walkers are indicated by the white circles except for $W_{0,1,2,3}$ which are specified by special markers. The preferred directions for the next time step are indicated by arrows. The ones of $W_{0,1,2,3}$ are reported as a function of $D_c$ in the lower panel. Three different regimes can clearly be identified.}
		\label{fig:pref_dir}
	\end{center}
\end{figure}

Let $W_0$ be the walker initially moved away from the stripe, and $W_{1,2,3}$ the three walkers neighboring the hole left by $W_0$, as shown in figure \ref{fig:pref_dir}. For each of these walkers, we identify which of their neighboring sites have the lowest concentration, which indicates their preferred direction for the next step. As shown in figure \ref{fig:pref_dir}, there are three regimes. First, for $D_c<0.255$, $W_1$, $W_2$ and $W_3$ would preferentially migrate to the same site. Because this is not possible, as the first one to jump will force the others to choose other sites, this creates a strong instability. No reformation of the original pattern is possible and the macroscopic structure is progressively lost. For $0.255<D_c<0.705$, the three walkers will preferentially migrate to different sites but the stripe will not be immediately formed back. This is a lightly unstable case since a small perturbation do not prevent the bands to reform but a larger defect can. Finally, for $D_c>0.701=D_c^*$, $W_{0,1,2,3}$ jump to 4 different sites in such a way that the original pattern is immediately recovered. This stability is the reason why the OS phase can emerge spontaneously from an initial random configuration as observed in our simulations. In fact, spontaneous fluctuations can form thin stripes locally, which will remain stable and be allowed to grow larger. However, if $D_c$ is too low, the small stripes that would emerge spontaneously could not grow as they would be unstable. 

\section{Summary and discussion}

In this paper, we have characterized the phase behavior of assemblies of auto-chemorepulsive walkers using a minimal lattice model mainly parametrized by the diffusion constant $D_c$ and the coupling of particles to the concentration field $\beta$. 

While alignment interactions are a common feature in many active systems, either due to dipole-dipole interactions or to particle shapes \cite{Baskaran2008, Solon2013, Ngo2014, Chatterjee2022}, the alignment observed in our model is of a very different nature because it is non-local in time and space and depends on the complete path followed by the walkers. As a result, alignment will occur only if walkers can travel persistently over a sufficiently long distance in order to produce a directed trail. This major difference with Vicsek-like and Ising-like models yields dissimilar band structures. Most strikingly, the chemo-repulsive walkers form bands which do not {\it a priori} have a size limit, whereas micro-phase separation with bands of a well-defined width is observed in most other band-forming active systems \cite{Marchetti2013, Joanny2010, Solon2015b, Chatterjee2023}. In addition, some sets of parameters can produce different steady-states, where a macroscopic band either contains all particles in the system or coexists with a gas or a dilute cloud traveling in the opposite direction. We also emphasize that we have observed similar bands by running simulations on a hexagonal lattice and by varying the aspect ratio of the simulation box. This indicates that the formation of bands is a robust feature of auto-chemorepulsive particles. 

In addition to their spatial structure, the behavior of the bands is remarkable. There is in fact no diffusion of walkers within the band as all walkers in the band travel ballistically at constant speed. If one of them takes a turn within the band, the fluctuation in the concentration field generated by this disturbance impacts neighbouring walkers which will in turn change their direction. This results in a cascading effect where a fluctuation of the concentration field travels within the band in the backward direction and eventually expel particles at the back of the band. This phenomenon is also observed as a band {\it absorbs} particles at its front. 

In addition to bands, more exotic structures are observed in our model that have rarely been found in other systems, namely the oscillating stripes. It is to this day not known whether the existence and stability of this phase is intrinsically due to the discrete nature of our lattice model, or whether it could be observed in off-lattice simulations. Investigating the details of this remarkable phase will motivate our future studies. 

Finally, we recall that the auto-chemorepulsive walk is a good example model for a non-markovian process with memory. Recent studies have been dedicated to quantifying the search efficiency of such walks \cite{Meyer2021, Barbier2022}, namely by calculating first-passage time properties. We emphasize here that, in this context, bands are very bad as they significantly increase the mean first-passage time for finding a target located at a random site of the lattice because walkers are densely distributed in a narrow band. While one might have expected that field-mediated repulsive interactions could be an efficient way to distribute walkers evenly in space and hence scan space rapidly, we show here that the emergence of such self-organized macroscopic structures are prohibitive for a reliable search. As a lead for future work, we will raise the question of possible strategies for the bands not to form and hence allow for better search strategies.

\section{Acknowledgments}
We acknowledge financial support by the DFG via the Collaborative Research Center SFB 1027.


%

\onecolumngrid

\end{document}